\documentclass[a4paper,11pt]{article}
\pdfoutput=1 
\usepackage[symbol]{footmisc} 
\usepackage{jheppub} 
	
\usepackage{color, colortbl}
\usepackage[T1]{fontenc}
\usepackage[utf8]{inputenc}
\usepackage[T1]{fontenc}
\usepackage{epsfig,latexsym}
\usepackage{amsmath}
\usepackage[dvipsnames]{xcolor}
\usepackage{adjustbox}
\usepackage{tensor}
\usepackage{graphicx}
\usepackage{verbatim}
\usepackage{mathrsfs}
\usepackage{amssymb}
\usepackage{soul}
\usepackage{multirow}
\usepackage{epsfig}
\usepackage{color,colordvi}
\usepackage{appendix}
\usepackage{slashed}
\usepackage{array}
\usepackage{cancel}
\usepackage{float}
\usepackage[font=footnotesize, justification=centering]{caption}
\usepackage{epsf}
\usepackage{amsmath}
\usepackage{bm}
\usepackage{mathtools}
\usepackage{physics}
\usepackage{rotating}
\usepackage{multirow}
\usepackage[normalem]{ulem}
\usepackage{tcolorbox}
\usepackage{bbold}
\usepackage{nicefrac}
\DeclareMathAlphabet{\mathpzc}{OT1}{pzc}{m}{it}

\newcommand{\g}{\sqrt{\abs{g}}}

\newcommand{\N}{\mathcal{N}}
\newcommand{\E}{\mathcal{E}}
\newcommand{\F}{\mathcal{F}}
\renewcommand{\H}{\mathcal{H}}
\renewcommand{\L}{\mathcal{L}}

\newcommand{\bchi}{\bar{\chi}}
\newcommand{\bpsi}{\bar{\psi}}
\newcommand{\bPsi}{\bar{\Psi}}
\newcommand{\btheta}{\bar{\theta}}

\newcommand{\lrnabla}{\overset{\leftrightarrow}{\nabla}}
\newcommand{\lnabla}{\overset{\leftarrow}{\nabla}}
\newcommand{\lrpar}{\overset{\leftrightarrow}{\partial}}

\newcommand{\fgamma}{\bar{\gamma}}
\newcommand{\mk}{\vec{k}^2}
\newcommand{\vk}{\vec{k}}
\newcommand{\vx}{\vec{x}}
\newcommand{\vfgamma}{\vec{\bar{\gamma}}}
\newcommand{\vpsi}{\vec{\psi}}

\newcommand{\kg}{\left(\vec{k}\cdot\vec{\fgamma}\right)}
\newcommand{\hO}{\hat{O}}

\newcommand{\vA}{\vec{A}}

\newcommand{\tA}{\tilde{A}}

\newcommand{\tP}{\tilde{\Pi}}

\newcommand{\cR}{C_\textup{R}}
\newcommand{\cI}{C_\textup{I}}
\newcommand{\dr}{d_\textup{R}}
\newcommand{\di}{d_\textup{I}}
\newcommand{\ta}{\tilde{a}}

\newcommand{\halpha}{\hat{\alpha}}
\newcommand{\hbeta}{\hat{\beta}}

\renewcommand{\[}{\left[}
\renewcommand{\]}{\right]}
\renewcommand{\(}{\left(}
\renewcommand{\)}{\right)}

\title{On energy and particle production in cosmology: 
the particular case of the gravitino}

\author[a]{Gabriele Casagrande,}
\author[a]{Emilian Dudas}
\author[b,c]{and Marco Peloso}

\affiliation[a]{Centre de Physique Th{\'e}orique, {\'E}cole Polytechnique, CNRS and IP Paris, 91128 Palaiseau Cedex, France 
} 
\affiliation[b]{Dipartimento di Fisica e Astronomia “Galileo Galilei”, Università di Padova, 35131 Padova, Italy}
\affiliation[c]{INFN, Sezione di Padova, 35131 Padova, Italy}

\emailAdd{gabriele.casagrande@polytechnique.edu, emilian.dudas@polytechnique.edu, marco.peloso@pd.infn.it}

\abstract{It is well-known that the number of particles produced in cosmology, commonly defined in the literature from the Fock space of the instantaneous hamiltonian of the canonically normalized fields, is ambiguous. On the other hand, the energy computed from the energy-momentum tensor should be physical. We compare the corresponding Fock spaces and relate them through a Bogolyubov transformation. We find that for particles of 
spin $0$, $1$ and $3/2$ the two Fock spaces are different, whereas they are the same for spin $1/2$ fermions. For spin $0$ and $1$, for particles of high momenta the two Fock spaces align, as intuitively expected. For the spin $3/2$, one finds two puzzles. The first one is that the two corresponding Fock spaces do not match even in the limit of high momenta. The second is that whereas we provide evidence for the equivalence theorem between longitudinal gravitinos and the goldstino in terms of an exact matching between the lagrangians and the instantaneous hamiltonians for the canonically normalized fields, the energy operator computed from the Rarita--Schwinger action does not seem to be captured in a simple way by the goldstino action. Our results suggest a re-analysis of non-thermal gravitino production in cosmology.}

\begin{document} 
\begin{flushright}
CPHT-RR064.102023
\end{flushright}

\maketitle

\flushbottom

\section{Introduction}\label{section:intro}

Particle production in cosmology is a major research subject bridging together quantum field theory, particle physics and general relativity methods. It is crucial in understanding particle creation during inflation and the consequent reheating of the universe \cite{parker,ford, Kofman:1997yn}. 
This phenomenon has been investigated also in the context of supergravity and applied in particular to the production of gravitinos in the early universe  \cite{Kallosh:1999jj, Kallosh:2000ve, Giudice:1999yt, Giudice:1999am, Nilles:2001fg, Nilles:2001ry, Hasegawa:2017hgd, Kolb:2021xfn, Kolb:2021nob,Dudas:2021njv}. The main motivation was to investigate whether the gravitino was compatible or not with primordial nucleosynthesis, testing in this way the realisation of the inflation paradigm in supergravity \cite{Kallosh:1999jj, Kallosh:2000ve, Giudice:1999yt, Giudice:1999am, Nilles:2001fg, Nilles:2001ry}. More recently, this last analysis was pushed to the point that the particle production mechanism was used as a probe of consistency of such constructions \cite{Hasegawa:2017hgd, Kolb:2021xfn, Kolb:2021nob,Dudas:2021njv}.

From its early days it was however realized that the concept of particle and therefore the number of produced particles is ambiguous in a time-dependent background \cite{books}. It is only well defined in the  asymptotic past and future, in the particular cases where time dependence is turned off, or it is negligible. On the other hand, backreaction of particle production on the spacetime geometry should be a physical, well-defined quantity, described by Einstein equations. Indeed, the matter energy-momentum tensor sources the metric and therefore particle production backreacts on the geometry, with  various potentially important consequences. 

The energy stored in the particle creation process is the most important quantity in these considerations. In the standard quantum field theory in flat space, the energy-momentum tensor is subject to some ambiguities leading to the so-called 
improved energy momentum tensors. However, the Noether charges, the hamiltonian and the 
total momentum are well-defined, and they can be computed via the standard procedures. Defined in this way, the energy and the canonical hamiltonian coincide and the latter is then used to quantise the system.
In time-dependent backgrounds, the notion of hamiltonian is instead subtle. 
The subtlety is partially related to time-dependent rescalings of the quantum fields\footnote{For the time being our discussion is general and the fields under consideration are of any spin.} $\Phi$, needed in order to put the kinetic terms into  a canonical form. The canonically-normalized fields $\varphi$ for a free theory have usually a simple time-evolution of the quantum 
oscillator type, governed by a hamiltonian $H_{\varphi} (\varphi,\Pi_{\varphi})$ obtained by  quantizing the lagrangian using the canonically normalized fields and their conjugate momenta $\Pi_{\varphi}$. We call this hamiltonian instantaneous in what follows.  Although this hamiltonian describes correctly the time evolution of $\varphi$, it fails to provide in most cases the correct energy operator, with the exception of spin $1/2$ fields. On the other hand, the correct energy operator is provided by the $00$ component of the energy-momentum
tensor $T^0{}_0$ computed starting from the original (non-canonically normalized) fields, which generally turns out to differ from the above hamiltonians.  The same result can be contained by working out the canonical hamiltonian using as variables the original (non rescaled) fields $\Phi$ of the original 
lagrangian. The corresponding energy operator correctly propagates in time the original fields $\Phi$, which have however a more complicated 
classical time-evolution. The mismatch between the two classes of hamiltonians can be anticipated by the fact that $\Phi$  have definite transformation properties under diffeomorphisms or Lorentz transformations, which
is not the case for the canonically normalized fields $\varphi$.  

From the field theoretical point of view, the difference between various hamiltonian formulations is well understood \cite{Fulling:1979ac, Weiss:1986gg, Bozza:2003pr, Grain:2019vnq}. However, the coupling with gravity promotes the energy-momentum tensor to a prominent physical role, so that also this difference between the field theoretical hamiltonians and the physical energy carries physical consequences and ambiguities once the quantisation of the system is performed. More specifically, it is the Fock space of the canonical hamiltonian $H_{\varphi} (\varphi,\Pi_{\varphi})$ which is commonly used in order to define the vacuum state $|  0  \rangle_{\varphi} $ and the particle excitations of momentum ${\bm k}$ of frequency $\omega_{\varphi}(k)$. This is done by introducing the creation 
and annihilation operators  $a_{\varphi}^{\dagger} ({\bm k})$,  $a_{\varphi}  ({\bm k})$, and Bogolyubov transformations of parameters $(\alpha_{\varphi} (k), \beta_{\varphi}(k)) $ which define the time-evolution of the Fock space of $H_{\varphi} (\varphi,\Pi_{\varphi})$. In this case, the number of produced particles of momentum ${\bm k}$ is typically $|\beta_{\varphi}(k)|^2$  and their corresponding "energy" is $ \omega_{\varphi}(k) |\beta_{\varphi}(k)|^2 $ \cite{Kofman:1997yn}. 
Since this hamiltonian typically differs from the genuine energy operator, the ground state defined through the canonical hamiltonian does not correspond to a minimum energy ground state. If one is interested in the physical  backreaction on the geometry of the 
quantum produced particles, one should instead quantize the energy operator $H_{\Phi} (\Phi,\Pi_{\Phi}) \sim T^0{}_0$ and build up its ground state of zero energy  $|  0  \rangle_{\Phi} $ and its Fock space via the creation and annihilation operators $a_{\Phi}^{\dagger} ({\bm k})$,  $a_{\Phi}  ({\bm k})$. Time evolution is implemented via a different set of Bogolyubov transformations  $(\alpha_{\Phi} (k), \beta_{\Phi}(k)) $.  Particle excitations in this case have a different frequency $\omega_{\Phi} (k)$, while the number of produced particles  is  $|\beta_{\Phi}(k)|^2$, having then energy   $\omega_{\Phi} (k) |\beta_{\Phi}(k)|^2$. The corresponding state does not 
have a definite number of particles in the Fock space of the alternative, more conventional, hamiltonian  $H_{\varphi} (\varphi,\Pi_{\varphi})$.  

The goal of the present paper is to work out explicitly the relation between the two different Fock spaces, which can be done analytically in the case of interest of an
FLRW expanding spacetime,  for the case of particles of spin $0,1/2,1$ and $3/2$.  The connection can be  described by a specific Bogolyubov transformation.  We check that in all cases, the two Fock spaces coincide in the limit of flat space and no time evolution, as it should.

Intuitively one would also expect the two Fock spaces to match also for very large momenta ${\bm k}$, where the effects of spacetime curvature are small. This holds indeed for the case of particles of spin
$0,1/2$ and $1$, but fails to be true for the Rarita--Schwinger field of spin $3/2$. This puzzling result has an interpretation for models of nonlinear, constrained supergravity \cite{Bergshoeff:2015tra, Ferrara:2015tyn}, like the ones discussed in \cite{Hasegawa:2017hgd, Kolb:2021xfn, Kolb:2021nob,Dudas:2021njv}, in terms of the UV cutoff which is naturally associated to such effective constructions and which forbids then to trust this UV limit. Because of this, it was also suggested in \cite{Kolb:2021xfn,Kolb:2021nob} that the effective supergravity theories that present such overproduction of longitudinal gravitinos should not belong to the landscape of quantum gravity but rather to the swampland.
However, the same behaviour occurs also in the case of linear supergravity coupled to one chiral multiplet \cite{Kallosh:1999jj, Kallosh:2000ve, Giudice:1999yt, Giudice:1999am, Nilles:2001fg, Nilles:2001ry}, theory which is in principle defined up to the Planck scale. We do not have a detailed understanding of this puzzling result, which definitely point out towards the need to use the Fock space
of the energy operator to discuss particle production and backreaction in this case. Actually, for all examples of fields of various spins we consider, the Bogolyubov transformation between the two Fock spaces does not converge fast enough to the identity in the UV. More precisely, the number of particles seen from  one  Fock space computed on the ground state of the other Fock space diverges. However, we believe that this should not be an obstacle in comparing energies in the two Fock spaces, since energy requires renormalization anyway. 

The physical significance of such difference between the two Fock spaces is supported also by the regularization and renormalization of the vacuum energy. Indeed, the zero-point energy subtraction from the energy operator must correspond to standard diffeomorphism-invariant counterterms in the original
lagrangian. This is crucial to discuss the backreaction of the produced particles and, more in general, for the consistency of the theory. We argue that, while the zero-point energy that one obtains from the ground state in the Fock space associated to the genuine energy operator $H_{\Phi} (\Phi,\Pi_{\Phi})$ can indeed be understood in this way, the one obtained from the zero-point energy of the ground state in the Fock space of the instantaneous  hamiltonian $H_{\varphi} (\varphi,\Pi_{\varphi})$ does not correspond to diffeomorphism-invariant counterterms. Since the field $\varphi$ is not a Lorentz scalar anyway, it is not obvious how to consistently renormalize the instantaneous hamiltonian $H_{\varphi} (\varphi,\Pi_{\varphi})$, which is  a procedure that would need to be performed if someone insisted in using the instantaneous hamiltonian to compute the physical energy density of the produced particles.

The particle production mechanism is also a good framework to test the validity of the equivalence theorem for the massive spin-$1$ and spin-$3/2$ fields. For the massive Proca spin-$1$ field, we find no subtlety for the equivalence theorem between the longitudinal component of the gauge field and the Goldstone boson, which is valid for large momenta $k \gg am$,  $k \gg H$. In the case of the massive Rarita--Schwinger spin $3/2$ field, we also investigate the equivalence theorem between the longitudinal component of the gravitino $\theta$ and the goldstino $G$ \cite{Fayet:1986zc, Casalbuoni:1988kv, Casalbuoni:1988qd}. We find that in the same limit of large momenta, an exact correspondence holds between the canonical lagrangians and instantaneous hamiltonians of the longitudinal gravitino in supergravity and the low-energy goldstino lagrangians with nonlinearly realised supersymmetry \cite{Volkov:1973ix, Komargodski:2009rz, Bonnefoy:2022rcw}.
We also compute in full details the energy-momentum tensor for
the original gravitino field in supergravity and its explicit expression in terms of the longitudinal gravitino in a FLRW background. The quantization of the corresponding $00$ component, which is the correct energy operator, provides
the appropriate framework for computing gravitino production backreaction on the geometry. However, we have no understanding of how to correctly identify this energy operator using directly the Goldstino lagrangians of \cite{Bonnefoy:2022rcw}, in the
decoupling limit of gravitational interactions. We leave this puzzle in the proof of the equivalence theorem for future work. 

The paper is organised as follows. In Sections \ref{section:spin_0}, \ref{section:spin_12}, \ref{section:spin_1} and \ref{section:spin_32} we discuss the above mentioned difference between the two Fock spaces for the fields of spin $0$, $1$ and $3/2$ respectively. We also discuss briefly the case of spin $1/2$ for completeness. Our presentation typically begins with the computation and quantisation of the instantaneous hamiltonian, proceeds with the calculation of the energy-momentum tensor and the quantisation of the correspondent energy operator and ends with the comparison between the two obtained Fock spaces. In Section \ref{section:norm_ord} we comment on the problem of the vacuum energy renormalization and its connection to the normal ordering of the ladder operators, while in Section \ref{section:eq_th} we discuss the  equivalence theorem for the spin-$3/2$ field. We finally conclude in Section \ref{sec:conclusion}. One  appendix contains some details of the decomposition of the spin $3/2$ Rarita--Schwinger Majorana fermion into longitudinal and transverse components. A second appendix provides details on the proof of the symmetry and conservation of the energy-momentum tensor for the Rarita--Schwinger field.

\section{The spin-0 field}\label{section:spin_0}


We begin our discussion from the real, free massive scalar field $\phi$ in a FLRW spacetime, which we parameterise in terms of the 
conformal time $\tau$ as
\begin{equation}
	g_{\mu\nu}=a^2(\tau) \ {}\eta_{\mu\nu} \ .
\label{eqn:flrw}
\end{equation}

\noindent The action for the field $\phi(\tau,\vec{x})$ is\footnote{We work with the signature $\left( +,-,-,- \right)$. A prime in what follows denotes derivative with respect to the conformal time $\tau$.}
\begin{equation}
\begin{aligned}
	S_\phi=&\int d^4x{}\g{} \ \mathcal{L}_\phi= \\
 =&\int d^4 x \g{} \ \frac{1}{2}\(\partial_\mu\phi\partial^\mu\phi-m^2\phi^2\)=\int d^4x{}\frac{a^2}{2}\[\(\phi'\)^2-\(\vec{\nabla}\phi\)^2-a^2m^2\phi^2\] \ .
\end{aligned}
\label{eqn:s_action}
\end{equation}

The kinetic term can be canonically normalised through the field redefinition
\begin{equation}
	\phi \equiv \frac{\varphi}{a} \ ,
\label{eqn:s_field_redef}
\end{equation}

\noindent leading to
\begin{equation}
\begin{aligned}
	S_\varphi=&\int d^4x \ {}\mathcal{L}_\varphi= \frac{1}{2} \int d^4x{} \left\{\(\varphi'\)^2-\varphi\(-\vec{\nabla}^2+a^2m^2-\frac{a''}{a}\)\varphi\right\} \ .
\end{aligned}
\label{eqn:s_cn_action}
\end{equation}

Working in momentum space via the decomposition\footnote{We denote with the subscript $\bm{k}$ the quantities that depend on the whole vector $\vec{k}$, while with $k$ those which depends only on its modulus. The $\cdot$ symbol denotes the euclidean scalar product $\bm{k}\cdot\bm{x}=\sum_i k_ix_i=-k_ix^i$.}
\begin{equation}
	\varphi\(\tau,\vec{x}\)=\int \frac{d^3k}{\(2\pi\)^{\nicefrac{3}{2}}}\left\{\varphi_{\bm{k}}(\tau)e^{i\bm{k}\cdot\bm{x}}+\bar{\varphi}_{\bm{k}}(\tau)e^{-i\bm{k}\cdot\bm{x}}\right\} \ ,
\end{equation}

\noindent where the bar symbol in $\bar{\varphi}_{\bm{k}}$ denotes complex conjugation, the equation of motion (EoM) takes the form of a harmonic oscillator:
\begin{equation}
	\varphi''_{\bm{k}}+\omega^2{}\varphi_{\bm{k}}=0 \ ,
\label{eqn:s_eom} 
\end{equation}

\noindent where
\begin{equation}
	\omega^2\equiv\vec{k}^2+a^2m^2-\frac{a''}{a}\equiv\omega_0^2-\frac{a''}{a},
 \label{eqn:omega}
\end{equation}

\noindent is the time-dependent frequency.
Notice that it is the canonically normalized field
$\varphi$ which has a simple, harmonic oscillator type field eq. (\ref{eqn:s_eom}), and not the original scalar field $\phi$. This makes $\varphi$ a natural variable for the canonical quantization, to which we now turn.

\subsection{Canonical quantisation and diagonalisation of the hamiltonian}

We now proceed with the computation and quantization of the hamiltonian. The hamiltonian density $\H_\varphi$ associated to the canonically normalised action (\ref{eqn:s_cn_action}) is given by
\begin{align}
\Pi_\varphi \equiv &\frac{\partial \L_\varphi}{\partial \varphi'}=\varphi',\\
\H_\varphi=&\frac{1}{2}\left\{\Pi_\varphi^2+\varphi\(-\vec{\nabla}^2+a^2m^2-\frac{a''}{a}\)\varphi\right\}.
\label{eqn:s_H_cn}
\end{align}

 \noindent Because of its relation to the time-dependent rescaling (\ref{eqn:s_field_redef}), we will refer to this hamiltonian (and to the analogue ones in the other spins cases) as instantaneous hamiltonian.~\footnote{A clearer terminology would be ``instantaneous hamiltonian {\it of the canonically normalized field}'', but we more simply denote it as ``instantaneous hamiltonian'' in this work for brevity.} 

We can now quantise this system by promoting the field $\varphi$ and its conjugate momentum $\Pi_\varphi$ to operators that satisfy the usual (equal-time) commutation relation
\begin{equation}
	\[\varphi(\tau,\vec{x}), \Pi_\varphi(\tau,\vec{y})\]=i{}\delta^{(3)}\(\vec{x}-\vec{y}\) \ . \label{eq:cr1}
\end{equation} 

\noindent This condition can be met by decomposing $\varphi$ and $\Pi_\varphi$ in terms of creation and annihilation operators as
\begin{equation}
\begin{aligned}
	\varphi(\tau,\vec{x})=&\int \frac{d^3k}{\(2\pi\)^{\nicefrac{3}{2}}}\left\{\varphi_{k}(\tau)e^{i\bm{k}\cdot\bm{x}}{}a_{\bm{k}}+\bar{\varphi}_k(\tau)e^{-i\bm{k}\cdot\bm{x}}{}a_{\bm{k}}^\dagger\right\},\\
	\Pi_\varphi(\tau,\vec{x})=&\int \frac{d^3k}{\(2\pi\)^{\nicefrac{3}{2}}}\left\{\Pi_k(\tau)e^{i\bm{k}\cdot\bm{x}}{}a_{\bm{k}}+\bar{\Pi}_k(\tau)e^{-i\bm{k}\cdot\bm{x}}{}a_{\bm{k}}^\dagger\right\}.
\end{aligned}
\label{eqn:s_q_dec}
\end{equation}

\noindent The mode functions $\varphi_k$ and $\Pi_k$ depend only on the magnitude of the momentum because the EoM (\ref{eqn:s_eom}) is isotropic. Imposing that also $a_{\bm{k}}$ and $a_{\bm{k}}^\dagger$ satisfy the usual commutation relation
\begin{equation}
	\[a_{\bm{k}},a_{\bm{k'}}^\dagger\]=\delta^{(3)}\(\vec{k}-\vec{k}'\), \qquad \qquad \[a_{\bm{k}},a_{\bm{k'}}\]=0=\[a_{\bm{k}}^\dagger,a_{\bm{k'}}^\dagger\],
\label{eqn:s_ca_comm}
\end{equation}

\noindent turns eq. (\ref{eq:cr1}) into the additional ("wronskian") condition
\begin{equation}
	\varphi_k\bar{\Pi}_k-\bar{\varphi}_k\Pi_k=i.
\label{eqn:s_wronsk}
\end{equation}

Making use of (\ref{eqn:s_q_dec}), the instantaneous hamiltonian (\ref{eqn:s_H_cn}) takes the form \cite{Nilles:2001fg}
\begin{equation}
\begin{aligned}
	H_\phi=&\int d^3k \ {}\H_\varphi=\frac{1}{2}\int d^3k {} \begin{pmatrix}a^\dagger_{\bm{k}} & a_{-\bm{k}}\end{pmatrix}\begin{pmatrix}\E_k & \F^\dagger_k\\ \F_k & \E_k\end{pmatrix}\begin{pmatrix}a_{\bm{k}} \\ a^\dagger_{-\bm{k}}\end{pmatrix} \ , 
\end{aligned}
\label{eqn:s_H_EF}
\end{equation}

\noindent with
\begin{align}
	\mathcal{E}_k=& |\Pi_k|^2+\omega^2|\varphi_k|^2 \ , & \mathcal{F}_k=\Pi_k^2+\omega^2\varphi_k^2 \ .
\label{eqn:s_E_F}
\end{align}

\noindent This hamiltonian can be diagonalized by means of a so-called Bogolyubov decomposition:
\begin{equation}
\begin{aligned}
    \varphi_k=&\frac{1}{\sqrt{2\omega}}\(\alpha_k+\beta_k\) \ , \\
	\Pi_k=&-\frac{i\omega}{\sqrt{2\omega}}\(-\alpha_k+\beta_k\) \ , 
\end{aligned}
\label{eqn:s_bog_dec}
\end{equation}

\noindent where $\alpha_k$ and $\beta_k$ are called Bogolyubov coefficients. The wronskian condition (\ref{eqn:s_wronsk}) turns into
\begin{equation}
    \abs{\alpha_k}^2-\abs{\beta_k}^2=1 \ ,
\end{equation}

\noindent while the EoM of the field $\varphi_k$ translate into
\begin{equation}
\left\{\begin{aligned}
	\alpha'_k=&-i\omega\alpha_k+\frac{\omega'}{2\omega}\beta_k \ , \\
	\beta'_k= & i\omega\beta_k +\frac{\omega'}{2\omega}\alpha_k \ .
\end{aligned}\right.{}  
\label{eqn:s_bog_eom}
\end{equation}

The Bogolyubov transformations are the set of transformations of the ladder operators which preserves the commutation relations (\ref{eqn:s_ca_comm}), namely mapping a set of ladder operators into a new, well-defined one:
\begin{equation}
	\begin{pmatrix} \tilde{a}_{\bm{k}}\\\tilde{a}_{-\bm{k}}^\dagger\end{pmatrix}=\begin{pmatrix} \alpha_k & \beta_k^\dagger \\ \beta_k & \alpha_k^\dagger\\\end{pmatrix}\begin{pmatrix} a_{\bm{k}}\\ a_{-\bm{k}}^\dagger\end{pmatrix} \ .
\label{eqn:s_bog_tr_1}
\end{equation}

\noindent After the transformation (\ref{eqn:s_bog_dec}), the instantaneous hamiltonian (\ref{eqn:s_H_EF}) becomes diagonal  \cite{Nilles:2001fg}
\begin{equation}
\begin{aligned}
	H_\varphi=&\frac{1}{2}\int d^3k{} \begin{pmatrix}a^\dagger_{\bm{k}} & a_{-\bm{k}}\end{pmatrix}\begin{pmatrix} \alpha_k^\dagger & \beta_k^\dagger \\ \beta_k & \alpha_k \\\end{pmatrix}\begin{pmatrix}\omega & 0\\ 0 & \omega\end{pmatrix}\begin{pmatrix} \alpha_k & \beta_k^\dagger \\ \beta_k & \alpha_k^\dagger\\\end{pmatrix}\begin{pmatrix}a_{\bm{k}} \\ a^\dagger_{-\bm{k}}\end{pmatrix}=\\
 =&\int d^3k{}\frac{\omega}{2}\(\tilde{a}_{\bm{k}}^\dagger\tilde{a}_{\bm{k}}+\tilde{a}_{\bm{k}}\tilde{a}_{\bm{k}}^\dagger\) \ ,
\end{aligned}
\label{eqn:s_H_diag}
\end{equation}
and the one of a harmonic oscillator of (time-dependent) frequency $\omega$. 

The number operator is defined as
$N_{\varphi} = \tilde{a}_{\bm{k}}^\dagger\tilde{a}_{\bm{k}}$ and the number of produced particles is given by its average on the Bunch--Davies (BD) vacuum defined by 
$a_{\bm{k}} | 0 \rangle =0$. One finds $\langle 0 | N_{\varphi} | 0 \rangle = |\beta_k|^2$, which implies that at the initial (conformal) time $\tau_0$ one has $\alpha_k (\tau_0)=1$, $\beta_k (\tau_0)=0$. The Bogolyubov transformation  (\ref{eqn:s_bog_tr_1}) becomes therefore trivial at the initial time. One can therefore define the BD vacuum as 
\begin{equation}
a_{\bm{k}} | 0 \rangle_{\rm BD} \equiv 0 \qquad   {\rm or }    \qquad {\tilde a}_{\bm{k}} (\tau_0)| 0 \rangle_{\rm BD} \equiv 0 \ . \label{eqn:BD}
\end{equation}

\noindent We remark that, defined in this way, the BD vacuum is the state that minimizes the instantaneous hamiltonian. We use this standard definition throughout the paper. We also note that the initial time $\tau_0$ should be chosen such that $\omega' \ll \omega^2$ so that the system remains close to the vacuum state for some time around $\tau_0$ (so that a change in the chosen $\tau_0$ produces negligible differences).

\subsection{The energy-momentum tensor and the physical energy}

The hamiltonian (\ref{eqn:s_H_cn}) correctly determines the EoM (\ref{eqn:s_eom}) and it is therefore the correct object to use in order to perform the quantisation of the field $\varphi$. However, it does not represent the physical energy of the system: this is encoded in the energy-momentum tensor
\begin{equation}
    T_{\mu\nu}\equiv\frac{2}{\g}\frac{\delta S}{\delta g^{\mu\nu}}=2\frac{\delta\mathcal{L}}{\delta g^{\mu\nu}}-g_{\mu\nu}\mathcal{L} \ ,
\label{eqn:emt_def}
\end{equation}  

\noindent which sources the Einstein equations. Going back to the starting (covariant) action (\ref{eqn:s_action}), the energy-momentum tensor of the scalar field results to be
\begin{equation}
	T_{\mu\nu}=\partial_\mu\phi\partial_\nu\phi-\frac{1}{2}g_{\mu\nu}\[\(\partial\phi\)^2-m^2\phi^2\] \ .
\label{eqn:s_T}
\end{equation}

The energy is then given by the $00$ component of this tensor, which is actually different from the instantaneous hamiltonian (\ref{eqn:s_H_cn}) once expressed in terms of the canonically normalised field: 
\begin{equation}
\begin{aligned}
	E_\varphi \equiv \int d^3x \g{} \ T^0{}_0=&\int d^3x{}\frac{a^2}{2}\left\{\(\phi'\)^2+\(\vec{\nabla}\phi\)^2+a^2m^2\phi^2\right\}=\\
 =&\frac{1}{2}\int d^3x{}\left\{\(\varphi'-aH\varphi\)^2+\varphi\(-\vec{\nabla}{}^2+a^2m^2\)\varphi\right\} \ .
\end{aligned}
\label{eqn:s_E}
\end{equation}

\noindent Performing the canonical quantisation (\ref{eqn:s_q_dec}) and the Bogolyubov decomposition (\ref{eqn:s_bog_dec}), one obtains
\begin{equation}
    E_\varphi=\frac{1}{2}\int d^3k{} \begin{pmatrix}\tilde{a}^\dagger_{\bm{k}} & \tilde{a}_{-\bm{k}}\end{pmatrix}\begin{pmatrix}\mathscr{E}_k & \mathscr{F}_k^\dagger\\ \mathscr{F}_k & \mathscr{E}_k\end{pmatrix}\begin{pmatrix}\tilde{a}_{\bm{k}} \\ \tilde{a}^\dagger_{-\bm{k}}\end{pmatrix} \ ,
\label{eqn:s_E_nd}
\end{equation}

\noindent where $\(\tilde{a}_{\bm{k}},\tilde{a}^\dagger_{\bm{k}}\)$ is the Fock space associated to the instantaneous hamiltonian (\ref{eqn:s_H_diag}), while the functions $\mathscr{E}_k$ and $\mathscr{F}_k$ are
\begin{equation}
\begin{aligned}
    \mathscr{E}_k=&\omega_0^2+\omega^2+a^2H^2 \ ,\\ \mathscr{F}_k=&\omega_0^2-\omega^2+a^2H^2-2iaH\omega \ .
\end{aligned}
\label{eqn:s_E_EF}
\end{equation}

Thus, a further Bogolyubov transformation is needed to diagonalize the energy operator (\ref{eqn:s_E_nd}):
\begin{equation}
	\begin{pmatrix} \tilde{a}_{\bm{k}} \\ \tilde{a}_{-\bm{k}}^\dagger\end{pmatrix}\equiv	\begin{pmatrix} \halpha_k & \hbeta_k^\dagger \\ \hbeta_k & \halpha_k^\dagger \end{pmatrix}\begin{pmatrix} \hat{a}_{\bm{k}} \\ \hat{a}_{-\bm{k}}^\dagger\end{pmatrix} \ ,
\label{eqn:s_2_bog}
\end{equation}

\noindent with 
\begin{equation}
	\halpha_k=\frac{\omega+\omega_0+iaH}{2\sqrt{\omega\omega_0}} \ , \qquad \qquad \hbeta_k=\frac{\omega-\omega_0-iaH}{2\sqrt{\omega\omega_0}} \ ,
\label{eqn:s_2_bog_coeff}
\end{equation}

\noindent where we recall that $\omega$ and $\omega_0$ differ according to eq. (\ref{eqn:omega}). The resulting energy is then
\begin{equation}
E_\varphi=\int d^3k{}\frac{\omega_0}{2}\(\hat{a}_{\bm{k}}^\dagger\hat{a}_{\bm{k}}+\hat{a}_{\bm{k}}\hat{a}_{\bm{k}}^\dagger\).
\label{eqn:s_E_diag}
\end{equation}

\subsection{Discussion on the two Fock spaces}

The diagonal Fock spaces associated to the instantaneous hamiltonian (\ref{eqn:s_bog_tr_1})-(\ref{eqn:s_H_diag}) and the energy operator (\ref{eqn:s_2_bog})-(\ref{eqn:s_E_diag}) are therefore different and the relation between the two is given by the Bogolyubov transformation (\ref{eqn:s_2_bog})-(\ref{eqn:s_2_bog_coeff}). 

According to the standard definition of the number operator in terms of the Fock space of the instantaneous hamiltonian (\ref{eqn:s_bog_tr_1})-(\ref{eqn:s_H_diag}), this mismatch between the two diagonal Fock spaces implies that the state with zero-particles does not have minimum energy. Indeed, using for convenience the second definition of the BD vacuum in eq. (\ref{eqn:BD}) one easily finds
\begin{equation}
_{\rm BD}\langle 0 | E_{\varphi} (\tau_0) | 0 \rangle_{\rm BD} =V_3 \[\omega_0 \(  \frac{1}{2}+ |{\hat \beta}_k|^2 \)\](\tau_0)  = V_3\[\frac{\omega^2 + \omega_0^2 + a^2 H^2}{4 \omega}\](\tau_0) \ ,  \label{eq:energyBD}     
\end{equation}

\noindent where we defined the space volume as the momentum space delta function
$V_3 = \delta^3 ({\bm k}=0)$.

In addition to the zero-point energy, there is therefore the non-vanishing contribution $\omega_0 |{\hat \beta}_k|^2$. The expression (\ref{eq:energyBD}) coincides with the energy of the BD vacuum only at the initial time $\tau_0$. One can also conversely check that the  minimum energy state does not have a definite number of particles. Actually, the total number of particles of the minimum energy states as seen from the Fock space of the instantaneous hamiltonian (\ref{eqn:s_H_diag}), obtained by integrating $|{\hat \beta}_k|^2$ over all momenta, is infinite. 

These results are a consequence of the curved spacetime setting. The genuine energy operator cannot be anything other than the object entering Einstein equations, the energy-momentum tensor. The hamiltonian is instead a differential operator which correctly provides the time evolution and, because of this, it is the correct object to use to perform its quantisation. The hamiltonian density $\H_\phi=\frac{\delta\L_\phi}{\delta\phi'}\phi'-\L_\phi$ and the $00$ component of the energy-momentum tensor do coincide as long as the original covariant formulation is used as a starting point. However, this formal identification is lost once time-dependent field redefinitions as (\ref{eqn:s_field_redef}) are performed: the time evolution of the new field is described by a new hamiltonian, here the instantaneous one (\ref{eqn:s_H_cn}), which is non trivially connected to the original one\footnote{From the point of view of the hamiltonian system, the allowed field redefinitions are canonical transformations, which preserve the structure of the Hamilton equations. When such transformations are time-dependent, the correspondent transformation of the hamiltonian is nontrivial. For a review, see \cite{Grain:2019vnq}.}. On the other hand, the correct energy operator $T^0{}_0$ (\ref{eqn:s_E}) differs from the instantaneous hamiltonian (\ref{eqn:s_H_cn}) once the time-dependent field redefinitions are applied. An appealing advantage of the instantaneous hamiltonian is that, contrary to $T^0{}_0$, it has the structure of that of a harmonic oscillator, which straightforwardly allows for a definition of an occupation number as in the case of constant frequency $\omega$. However, as we argued, its use for the computation of the produced energy is less straightforward.

Nevertheless, the two Fock spaces are expected to agree at least in the UV limit $k\to \infty$, in which the gravitational effects should be suppressed. This condition is fulfilled if
\begin{equation}
    \lim_{k\to\infty}\hat{\beta}_k=0
\label{eqn:s_uv_lim}
\end{equation}

\noindent is satisfied. From eq. (\ref{eqn:s_2_bog_coeff}) one immediately sees that this is indeed the case for the scalar field, since for $k\to\infty$ one has $\omega\to\omega_0$.
As already mentioned, since $\hat{\beta}_k \sim 1/k$ for large momenta,  the total number of particles in one Fock space computed on the ground state of the other Fock space diverges. 
This divergence, in and of itself, is not an insurmountable problem, as one should rather compute the produced energy, which requires renormalization. We discuss this issue in Section \ref{section:norm_ord}, where we argue that also renormalization is more easily performed starting from the energy operator $T^0_0$ rather than from the instantaneous hamiltonian.

\section{The spin-\texorpdfstring{${\nicefrac{\bm 1}{\bm 2}}$}{1/2} field}\label{section:spin_12}

Although there is no subtlety in this case, we also briefly summarize the case of the  spin-$\nicefrac{1}{2}$ Dirac fermion $\psi$. The curved spacetime Dirac matrices $\gamma^\mu$ are defined in terms of the flat one $\fgamma^a$ as
\begin{align}
	\gamma^\mu \equiv e^\mu_a\,\bar{\gamma}^a \ , && e^\mu_a=a^{-1}\delta^\mu_a \ ,
\end{align}

\noindent with $e^\mu_a=\(e^a_\mu\)^{-1}$ being the veirbein associated to the FLRW metric (\ref{eqn:flrw}). The starting action is
\begin{equation}
	S_\psi=\int d^4x{} \ e{} \ \mathcal{L}_\psi=\int d^4x {}{} \ e{}{} \ \bar{\psi}\[\frac{i}{2}\gamma^\mu\lrnabla_\mu-m\]\psi \ ,
\label{eqn:f_action}
\end{equation}

\noindent where $e\equiv\text{det}\(e^a{}_\mu\)$ and the kinetic term is written in compact notation for
\begin{equation}
     \bpsi\gamma^\mu\lrnabla_\mu\psi\equiv \bpsi\gamma^\mu\nabla_\mu\psi -\bpsi\overset{\leftarrow}{\nabla}{}_\mu\gamma^\mu\psi,
\end{equation}

\noindent with
\begin{align}
    \nabla_\mu\psi \equiv \partial_\mu\psi +\frac{1}{4}\(\omega_\mu\)_{ab}\fgamma^{ab}\psi,  && 
   \bpsi\overset{\leftarrow}{\nabla}_\mu \equiv \partial_\mu\bpsi -\frac{1}{4}\(\omega_\mu\)_{ab}\bpsi\fgamma^{ab}.
\label{eqn:f_der_def}
\end{align}

\noindent The corresponding Dirac EoM is
\begin{equation}
    \(i\gamma^\mu\nabla_\mu-m\)\psi=0 \ .
\label{eqn:f_eom}
\end{equation} 

The spin-1/2 kinetic term can be canonically normalised by the time-dependent rescaling
\begin{equation}
	\psi \equiv a^{-\nicefrac{3}{2}}\,\chi \ ,
\label{eqn:f_fr}
\end{equation}
\noindent and the associated action becomes
\begin{equation}
	S_\chi=\int d^4x{} \ \L_\chi=\int d^4x{} \ \bchi\[\frac{i}{2}\fgamma^\mu\lrpar_\mu-am\]\chi \ ,
\label{eqn:f_action_cn}
\end{equation}

\noindent which is identical to the case of a Dirac fermion propagating in a flat spacetime with a time-dependent mass $a m$.

We can now proceed with the computation of the instantaneous hamiltonian. The conjugate momenta are
\begin{align}
	{\Pi}^\dagger_\chi\equiv&\frac{\delta\mathcal{L}_\chi}{\delta\partial_0\chi}=\frac{i}{2}\chi^\dagger \ , & \Pi_{\chi}\equiv&\frac{\delta\mathcal{L}_\chi}{\delta\partial_0\chi^\dagger}=-\frac{i}{2}\chi \ ,
\label{eqn:f_Pi_cn}
\end{align}

\noindent and the hamiltonian density is computed to be
\begin{equation}
\begin{aligned}
	H_\chi=&\int d^3x\({\Pi}_\chi\partial_0\chi+\partial_0\chi^\dagger\Pi_{\chi}-\mathcal{L}_\chi\)=\int d^3x\(-\frac{i}{2}\bchi\fgamma^i\lrpar_i\chi+am\bchi\chi\) \ ,
\end{aligned}
\label{eqn:f_H_cn}
\end{equation}

\noindent where the symbol $\lrpar_\mu$ is defined in analogy to (\ref{eqn:f_der_def}).

This result has to be compared with the energy-momentum tensor, which for a fermion is defined as
\begin{equation}
	T^a_\mu\equiv\frac{1}{e}\frac{\delta S}{\delta e^\mu_a} \ .
\label{eqn:emt_def_e}
\end{equation}

\noindent On-shell, for the action  (\ref{eqn:f_action}) this tensor is computed to be\footnote{Symmetrisation and antisymmetrisation of indices is denoted respectively with round and squared parenthesis $()$ and $[]$ and it is always defined with weight one.}
\begin{equation}
	T^{\mu\nu}=\frac{i}{2}\bpsi\gamma^{(\mu}\lrnabla\,^{\nu)}\psi,
\label{eqn:f_T}
\end{equation}

\noindent so that the energy operator is given by
\begin{equation}
\begin{aligned}
	E_\chi=&\int d^3x{} \ e{} \ T^0{}_0=\int d^3x \ a^4\(\frac{i}{2}\bpsi\gamma^0\lrnabla_0\psi\)\overset{(\ref{eqn:f_eom})}{=}\\
	=&\int d^3x \,a^4\[-\frac{i}{2}\bpsi\gamma^i\lrpar_i\psi+m\bpsi\psi\]\overset{(\ref{eqn:f_fr})}{=}\int d^3x\(-\frac{i}{2}\bchi\fgamma^i\lrpar_i\chi+am\bchi\chi\).
\end{aligned}
\label{eqn:f_E}
\end{equation}

Therefore, the instantaneous hamiltonian and the energy operator coincide for the Dirac spin-$\nicefrac{1}{2}$ field, which means that there is no difference between the two respective diagonal Fock spaces of the quantised theory, which is discussed for example in \cite{Nilles:2001fg, Peloso:2000hy, Chung:2011ck}. 

\section{The spin-1 field}\label{section:spin_1}

We proceed in our analysis with the case of a massive Proca gauge boson, for which we refer the reader to \cite{Himmetoglu:2008zp, Graham:2015rva, Ahmed:2020fhc} for more details. The starting action is
\begin{equation}
	S_A=\int d^4x\,\g\left\{-\frac{1}{4}F^{\mu\nu}F_{\mu\nu}+\frac{M^2}{2}A^\mu A_\mu\right\} \ ,
\label{eqn:b_action_gen}
\end{equation}

\noindent which in a FLRW background takes the form
\begin{equation}
\begin{aligned}
	S_A=&\int d^4x\,\left\{-\frac{1}{4}F^{\mu\nu}F_{\mu\nu}+\frac{a^2M^2}{2}A^\mu A_\mu\right\}=\\
	=&\int d^4x\,\left\{-\frac{1}{2}F^{0i}F_{0i}-\frac{1}{4}F^{ij}F_{ij}+\frac{a^2M^2}{2}\left(A^0A_0+A^iA_i\right)\right\} \ ,
\end{aligned}
\label{eqn:b_action_op}
\end{equation}

\noindent where in the last line the indices are contracted with the flat metric $\eta_{\mu\nu}$. 

The $A_0$ component is actually not a dynamical field since its EoM is a constraint:
\begin{equation}
	\partial_0\partial_i A^i-\partial_i\partial^iA_0-a^2M^2A_0=0 \ .
	\label{eqn:b_A0_const}
\end{equation}

\noindent This constraint can be easily solved going to momentum space via
\begin{align}
	A_\mu(\tau,\vec{x})=\int \frac{d^3k}{(2\pi)^{\nicefrac{3}{2}}}A_{\mu,\bm{k}}(\tau)e^{i\bm{k}\cdot\bm{x}} \ , && \bar{A}_{\mu,\bm{k}}=A_{\mu,-\bm{k}} \ .
\label{eqn:b_mom_sp}
\end{align}

\noindent One can decompose $A_i$ into a longitudinal and a transverse part according to
\begin{equation}
	A_{i,\bm{k}}=-ik_i A_\textup{L}+A_\textup{T}{}_{i,\bm{k}} \ ,
\label{eqn:b_lt_dec}
\end{equation}
\noindent with
\begin{align}
	k^i A_\textup{T}{}_{i,\bm{k}}=&0 \ , & A_\textup{L}&=-i\frac{k_i A^i_{\bm{k}}}{\vk^2} \ ,  & \bar{A}_{i,\bm{k}}=ik_i\bar{A}_\textup{L}+\bar{A}_\textup{T}{}_{i,\bm{k}} \ ,
\end{align}
\noindent so that the constraint equation (\ref{eqn:b_A0_const}) is solved by
\begin{equation}
	A_{0,\bm{k}}=\frac{\vk^2}{\vk^2+a^2M^2}A'_\textup{L} \ .
\label{eqn:b_A0_const_k}
\end{equation}

\noindent Then, the action (\ref{eqn:b_action_op}) splits into two decoupled sectors:
\begin{align}
	S=&\int d\tau d^3x \left\{\mathcal{L}_\textup{T}+\mathcal{L}_\textup{L}\right\} \ , \label{eqn:b_action_lt}\\
	\mathcal{L}_\textup{T}=&\frac{1}{2}\abs{\vA{}'_\textup{T}}^2-\frac{\omega_0^2}{2}\abs{\vA_\textup{T}}^2 \ , \label{eqn:b_action_t}\\
	\mathcal{L}_\textup{L}=&\frac{\vk^2}{2}\[N^2\abs{A'_\textup{L}}^2-a^2M^2\abs{A_\textup{L}}^2\] \ , \label{eqn:b_action_l}
\end{align}

\noindent where
\begin{align}
	\omega_0^2 \equiv & \, \vec{k}{}^2+a^2M^2 \qquad ,  & N^2\equiv \frac{a^2M^2}{\omega_0^2} \ . 
\end{align}

The kinetic term of the longitudinal field $A_\textup{L}$ can be canonically normalised by means of the redefinition
\begin{equation}
	A_\textup{L}=\frac{1}{\abs{\vk}N}\tilde{A} \ ,
\label{eqn:b_cn_redef}
\end{equation}

\noindent so that the lagrangian (\ref{eqn:b_action_l}) can finally be written in the form of that of a harmonic oscillator:
\begin{equation}
	\mathcal{L}_\textup{L}=\frac{1}{2}\(\abs{\tA'}^2-\omega^2\abs{\tA}^2\) \ ,
 \label{eqn:b_lag_tA}
\end{equation}

\noindent with
\begin{align}
	\omega^2\equiv & \ \omega_0^2-\frac{N''}{N} \ , & \frac{N''}{N} = \ &\frac{\vk^2}{\vk^2+a^2M^2}\(\frac{a''}{a}-3N^2a^2H^2\) \ .
\label{eqn:b_w}
\end{align}

Note that, written in this form, the action for the longitudinal component of the massive gauge boson resembles the one for the real scalar field we discussed in Section \ref{section:spin_0}, via the replacement 
\begin{align}
	\frac{a'}{a} & \longleftrightarrow \frac{N'}{N} \ , & \frac{a''}{a} & \longleftrightarrow \frac{N''}{N} \ .
\label{eqn:b_replace_rules}
\end{align}
Notice that for large momenta the correspondences in (\ref{eqn:b_replace_rules}) become equalities. This fact will allow for an immediate connection with the equivalence theorem.
Notice the well-known and important
fact that in the small mass limit the time-dependence and therefore the production of transverse polarizations of the gauge bosons is suppressed. On the other hand, the production of the longitudinal components is not, in analogy with the case of a massless scalar. 
Longitudinal components will be therefore copiously produced during inflation; in particular, if stable enough, a significant amount of vector dark matter can be produced gravitationally \cite{ Graham:2015rva, Ahmed:2020fhc}. 

\subsection{Canonical quantisation and diagonalisation of the hamiltonian}

The instantaneous hamiltonian associated to the canonically normalised, longitudinal component of the massive gauge boson is given by\footnote{We denote the complex conjugate of the field $\tA$ with $\tA^*$.}
\begin{align}
    \tilde{\Pi}\equiv &\frac{\partial \L_\textup{L}}{\partial \tA'}=\frac{\tA^*{}'}{2} \ , \label{eqn:b_cm}\\
    H_\textup{L}=&\int d^3k\H_\textup{L}=\int d^3k\frac{1}{2}\(\abs{\tP}^2+\omega^2\abs{\tA}^2\) \ . \label{eqn:b_H}
\end{align}

The quantization of the system proceeds then promoting $\tA$ and $\tP$ to operators satisfying the standard commutation relation in position space:
\begin{equation}
	\[\tA(\tau,\vx),\tP(\tau,\vec{y})\]=i\delta^{(3)}\(\vx-\vec{y}\) \ .
\label{eqn:b_comm_rel_xy}
\end{equation}

\noindent This condition allows for a decomposition in momentum space, compatible with (\ref{eqn:b_mom_sp}), in terms of ladder operators:
\begin{align}
	\tA(\tau,\vk)=&\lambda_k a_{\bm{k}}+\bar{\lambda}_k a_{-\bm{k}}^\dagger \ , & \tP(\tau,\vk)=&\pi_k a_{\bm{k}}+\bar{\pi}_k a_{-\bm{k}}^\dagger \ ,
\label{eqn:b_tA_lad}
\end{align}

\noindent where the mode functions $\lambda_k$ and $\pi_k$ depend only on the modulus $k$ of the momentum because the EoM of the field $\tA$ are isotropic. The ladder operators satisfy the canonical commutation relations
\begin{align}
	\[a_{\bm{k}},a_{\bm{q}}^\dagger\]=&\delta^{(3)}\(\vk-\vec{q}\) \ , & \[a_{\bm{k}},a_{\bm{q}}\]=&0=\[a_{\bm{k}}^\dagger,a_{\bm{q}}^\dagger\] \ ,
\label{eqn:b_lad_op_comm}
\end{align}

\noindent if one imposes the additional wronskian condition on $\varphi_k$ and $\pi_k$:
\begin{equation}
	\lambda_k\bar{\pi}_k-\bar{\lambda}_k {\pi}_k=i \ .
\end{equation}

\noindent The instantaneous hamiltonian (\ref{eqn:b_H}) becomes then
\begin{equation}
\begin{aligned}
	H_\textup{L}=&\int d^3k\frac{1}{2}\begin{pmatrix} a_{\bm{k}}^\dagger & a_{-\bm{k}}\end{pmatrix}\begin{pmatrix} \mathcal{E}_k &\mathcal{F}_k^\dagger\\ \mathcal{F}_k & \mathcal{E}_k\end{pmatrix}\begin{pmatrix} a_{\bm{k}} \\ a_{-\bm{k}}^\dagger\end{pmatrix} \ ,
\end{aligned}
\label{eqn:b_H_EF}
\end{equation}

\noindent with
\begin{align}
	\mathcal{E}_k=& |\pi_k|^2+\omega_0^2|\lambda_k|^2 \ , & \mathcal{F}_k=\pi_k^2+\omega_0^2\lambda_k^2 \ .
\label{eqn:b_E_F}
\end{align}

We observe that the hamiltonian (\ref{eqn:b_H_EF})  exactly matches the one for the real scalar field (\ref{eqn:s_H_EF}) through the replacement rule (\ref{eqn:b_replace_rules}). This suggest to use the analogue of the Bogolyubov transformation (\ref{eqn:s_bog_dec}) to diagonalize the hamiltonian (\ref{eqn:b_H_EF}). We thus decompose the mode functions $\lambda_k$ and $\Phi_k$ as
\begin{equation}
\begin{aligned}
    \lambda_k=&\frac{1}{\sqrt{2\omega}}\(\alpha_k+\beta_k\) \ , \\
	\pi_k=&-\frac{i\omega}{\sqrt{2\omega}}\(-\alpha_k+\beta_k\) \ , 
\end{aligned}
\label{eqn:b_bog_dec}
\end{equation}

\noindent with $\alpha_k$ and $\beta_k$ satisfying a system of equations of motion analogue to (\ref{eqn:s_bog_eom}). Defining
\begin{equation}
	\begin{pmatrix} \tilde{a}_{\bm{k}}\\\tilde{a}_{-\bm{k}}^\dagger\end{pmatrix}=\begin{pmatrix} \alpha_k & \beta_k^\dagger \\ \beta_k & \alpha_k^\dagger\\\end{pmatrix}\begin{pmatrix} a_{\bm{k}}\\ a_{-\bm{k}}^\dagger\end{pmatrix} \ ,
\label{eqn:b_bog_tr_1}
\end{equation}

\noindent we obtain the instantaneous hamiltonian in diagonal form:
\begin{equation}
	H_\textup{L}=\int d^3k{}\frac{\omega}{2}\(\tilde{a}_{\bm{k}}^\dagger\tilde{a}_{\bm{k}}+\tilde{a}_{\bm{k}}\tilde{a}_{\bm{k}}^\dagger\) \ .
\label{eqn:b_H_diag}
\end{equation}

\noindent We note again that the diagonalisation of the hamiltonian for the longitudinal gauge field is completely equivalent to the one of the real scalar field discussed in Section \ref{section:spin_0}.

\subsection{The energy-momentum tensor}

We now turn to the energy operator. The energy-momentum tensor associated to the starting action (\ref{eqn:b_action_gen}) is 
\begin{equation}
	T^{\mu\nu}=-F^{\mu\alpha}F^\nu_\alpha+\frac{1}{4}g^{\mu\nu}\left(F^{\alpha\beta}F_{\alpha\beta}\right)+M^2\left(A^\mu A^\nu-\frac{1}{2}g^{\mu\nu}A^\alpha A_\alpha\right) \ .
\label{eqn:b_emt}
\end{equation}

\noindent The energy density is then given by
\begin{equation}
	E_A=\int d^3x \g \ T^0{}_0=\int d^3x\,\[-\frac{1}{2}F^{0i}F_{0i}+\frac{1}{4}F^{ij}F_{ij}+\frac{a^2M^2}{2}\left(A^0A_0-A^iA_i\right)\] \ .
	\label{eqn:b_H_EE}
\end{equation}

Going to momentum space via (\ref{eqn:b_mom_sp}) and canonically normalising the longitudinal field via (\ref{eqn:b_cn_redef}), the energy density decomposes as\footnote{The quantity $\vec{\Pi}_\textup{T}$ is the conjugate momentum associated to $\vec{A}_\textup{T}$ and can be computed from lagrangian (\ref{eqn:b_action_t}) to be $\vec{\Pi}_\textup{T}^*=\frac{\vec{A}_\textup{T}'}{2}$.}
\begin{align}
	E_A=&\int d^3k \(\mathcal{E}_\textup{T}+\mathcal{E}_\textup{L}\) \ , \label{eqn:b_H_EE_lt}\\
	\mathcal{E}_\textup{T}=&\frac{1}{2}\abs{\vec{\Pi}_\textup{T}}^2+\frac{\omega_0^2}{2}\abs{\vec{A}_\textup{T}}^2 \ , \label{eqn:b_E_t}\\
	\mathcal{E}_\textup{L}=&\frac{1}{2}\abs{\tP-\frac{N'}{N}\tA}^2+\frac{\omega_0^2}{2}\abs{\tA}^2 \ .
\end{align}

\noindent Quantising the theory as in (\ref{eqn:b_tA_lad}) and after applying the Bogolyubov decomposition (\ref{eqn:b_bog_dec}), the energy density for the longitudinal component takes the form
\begin{equation}
	\mathcal{E}_\textup{L}=\frac{1}{4\omega}\begin{pmatrix} \tilde{a}_{\bm{k}}^\dagger & \tilde{a}_{-\bm{k}}\end{pmatrix}\begin{pmatrix}\mathscr{E}_k & \mathscr{F}_k^\dagger\\ \mathscr{F}_k & \mathscr{E}_k\end{pmatrix}\begin{pmatrix} \tilde{a}_{\bm{k}} \\ \tilde{a}_{-\bm{k}}^\dagger\end{pmatrix} \ ,
\label{eqn:b_E_EF_bog}
\end{equation}

\noindent with
\begin{align}
		\mathscr{E}_k=&\omega_0^2+\omega^2+\(\frac{N'}{N}\)^2 \ ,  &
		\mathscr{F}_k=&\omega_0^2-\omega^2+\(\frac{N'}{N}\)^2+2i\frac{N'}{N}\omega \ . \label{eqn:b_EF_bog}
\end{align}

Therefore, as for the real scalar field, in order to diagonalize the energy (\ref{eqn:b_E_EF_bog}) we need a further Bogolyubov transformation, given by
\begin{equation}
	\begin{pmatrix} \tilde{a}_{\bm{k}} \\ \tilde{a}_{-\bm{k}}^\dagger\end{pmatrix}\equiv	\begin{pmatrix} \halpha_k & \hbeta_k^\dagger \\ \hbeta_k & \halpha_k^\dagger \end{pmatrix}\begin{pmatrix} \hat{a}_{\bm{k}} \\ \hat{a}_{-\bm{k}}^\dagger\end{pmatrix} \ ,
\label{eqn:b_diag_lad}
\end{equation}

\noindent with 
\begin{equation}
	\hat{\alpha}=\frac{1}{2\sqrt{\omega\omega_0}}\(\omega+\omega_0+i\frac{N'}{N}\) \ , \qquad \qquad \hat{\beta}=\frac{1}{2\sqrt{\omega\omega_0}}\(\omega-\omega_0-i\frac{N'}{N}\) \ .
\label{eqn:b_ab_sol}
\end{equation}

\noindent The diagonal energy operator is then
\begin{equation}
    E_\textup{L}=\int d^3k\frac{\omega_0}{2}\(\hat{a}_{\bm{k}}^\dagger\hat{a}_{\bm{k}}+\hat{a}_{\bm{k}}\hat{a}_{\bm{k}}^\dagger\) \ .
\label{eqn:b_diga_E}
\end{equation}

\subsection{Discussion on the two Fock spaces and the equivalence theorem}

We found out that also for the massive gauge boson the Fock spaces in which the instantaneous hamiltonian and the energy operator have a diagonal form are different. This difference is parameterised by the Bogolyubov transformation (\ref{eqn:b_diag_lad})-(\ref{eqn:b_ab_sol}). The two Fock spaces do coincide in the UV limit $k\to\infty$: like for the scalar field, in this limit one has $\omega\to\omega_0$, so that
\begin{equation}
    \lim_{k\to\infty}\hat{\beta}_k=0 \ .
\end{equation}

The same comments made around  (\ref{eq:energyBD}) for the scalar field apply also in this case: the BD vacuum does not have a minimum energy at the initial time, and it is given by
\begin{equation}
_{\rm BD}\langle 0 | E_L (\tau_0)| 0 \rangle_{\rm BD} = V_3\[\omega_0 \(  \frac{1}{2}+ |{\hat \beta}_k|^2 \)\](\tau_0) = V_3\[\frac{\omega^2 + \omega_0^2 + \(\nicefrac{N'}{N}\)^2}{4 \omega}\] (\tau_0) \ . \label{eq:energyprocaBD}     
\end{equation}

\noindent Moreover, $\hat{\beta}_k \sim 1/k$ for large momenta and the total number of particles in one Fock space computed on the ground state of the other Fock space diverges, when integrating over momenta. The comparison of the energies of the ground states in both Fock spaces should however be meaningful,  since energy requires in any case renormalization.

As we already noticed several times, the analysis of the longitudinal mode of the massive gauge boson follows step by step the one of the real scalar field. This correspondence can be made more quantitative through the equivalence theorem, which states that the amplitudes associated to the longitudinal polarisation of the massive gauge boson theory are equivalent to the amplitudes of a real scalar field $z$ (the would-be Goldstone boson associated to the gauge symmetry breaking) via the substitution 
\begin{equation}
	A_\mu  \longrightarrow  \frac{1}{M}\partial_\mu z \ .
\label{eqn:b_et_repl}
\end{equation}

\noindent Via this replacement, the starting action (\ref{eqn:b_action_gen}) becomes the one for a free scalar field:
\begin{equation}
	S_A=\int d^4x\g\[-\frac{1}{4}F^{\mu\nu}F_{\mu\nu}+\frac{M^2}{2}A^\mu A_\mu\]  \longrightarrow  \int d^4x \g \ \frac{1}{2}\partial_\mu z \partial^\mu z \ .
\end{equation}

The equivalence theorem works for energies much higher than the mass of the gauge boson, i.e. for $\vk^2\gg a^2 M^2$. In this limit, the correspondence (\ref{eqn:b_replace_rules}) is actually realised manifestly:
\begin{align}
\frac{N'}{N}&=\frac{\vk^2}{\vk^2+a^2M^2}aH=aH+\mathcal{O}\(\frac{a^2M^2}{\vk^2}\) \ , \\
\frac{N''}{N}&=\frac{\vk^2}{\vk^2+a^2M^2}\[\frac{a''}{a}-\frac{3M^2a^4H^2}{\vk^2+a^2M^2}\]=\frac{a''}{a}+\mathcal{O}\(\frac{a^2M^2}{\vk^2}\) \ . 
\end{align}

\noindent Therefore, the equivalence theorem is the underlying reason of this correspondence in the UV limit.

\subsection{The transverse gauge field}

We conclude the section on the massive spin-1 gauge field by commenting on the Fock spaces associated to the transverse components. One can immediately compute the instantaneous hamiltonian from the lagrangian (\ref{eqn:b_action_t}), obtaining
\begin{equation}
    \H_\textup{T}=\frac{1}{2}\abs{\vec{\Pi}_\textup{T}}^2+\frac{\omega_0^2}{2}\abs{\vec{A}_\textup{T}}^2.
\end{equation}

\noindent We see therefore that the instantaneous hamiltonian coincides with the energy operator, computed in (\ref{eqn:b_E_t}), so that, as for the spin-$\nicefrac{1}{2}$ field, there are no ambiguities related to different Fock spaces.

\section{The spin-\texorpdfstring{$\nicefrac{\bm 3}{\bm2}$}{3/2} field}\label{section:spin_32}

Finally, we turn  to the spin-$\nicefrac{3}{2}$ field, which was our main motivation for this work and which, as one will see, features some interesting puzzles (at least for us). The action we are considering is the massive Rarita--Schwinger action coupled to a real scalar field\footnote{The Dirac matrices combinations $\gamma^{\mu\nu}$ and $\gamma^{\mu\nu\rho}$ appearing in (\ref{eqn:L_psi}) are defined respectively as $\gamma^{\mu\nu}\equiv\gamma^{[\mu}\gamma^{\nu]}$ and $\gamma^{\mu\nu\rho}\equiv\gamma^{[\mu}\gamma^\nu\gamma^{\rho]}=\frac{1}{2}\(\gamma^\mu\gamma^{\nu\rho}+\gamma^{\nu\rho}\gamma^\mu\)$.}
\begin{equation}
\begin{aligned}
	S_\Psi=&\int d^4x \ e \ \mathcal{L}_\Psi=\\
 =&\int d^4x \ e\[-\frac{1}{2}\(i\bPsi_\mu\gamma^{\mu\nu\rho}\nabla_\nu\Psi_\rho+m(\phi)\bPsi_\mu\gamma^{\mu\nu}\Psi_\nu\)+\frac{1}{2}\partial_\mu\phi\partial^\mu\phi-V(\phi)\] \ , 
\end{aligned}
\label{eqn:L_psi}
\end{equation}

\noindent where the spin-$\nicefrac{3}{2}$ field $\Psi_\mu$ is taken to be a Majorana fermion and the mass parameter $m$ is assumed to be field-dependent, which is generically the case in supergravity. This theory is in fact a subsector of many $\N=1$, $D=4$ supergravity constructions with spontaneously broken supersymmetry. This is the case especially of constrained, nonlinear realisations \cite{Bergshoeff:2015tra, Ferrara:2015tyn}, but also of usual supergravity coupled to one chiral multiplet\footnote{In this last case, the chiral multiplet is the one of the inflaton.The inflatino is the goldstino during inflation, which can be gauged away in the unitary gauge. The scalar is complex, so in addition to the inflaton, there is an additional real scalar in the spectrum. Our current setup applies if the additional scalar is very heavy or frozen during inflation. If the dynamics of the additional scalar is non-trivial, some changes are expected, that could maybe explain some of our results.} \cite{Kallosh:1999jj, Kallosh:2000ve}.

\subsection{General equations and the FLRW spacetime}

The gravitino EoM are
\begin{equation}
	\gamma^{\mu\nu\rho}\nabla_\nu\Psi_\rho=im\gamma^{\mu\nu}\Psi_\nu \ . 
\label{eqn:gr_eom_1}
\end{equation}

\noindent Two constraints are found by contracting this equation with $\gamma^\mu$ and $\nabla_\mu$ respectively:
\begin{align}
    \gamma^{\mu\nu}\nabla_\mu\Psi_\nu=&\frac{3}{2}im\gamma^\mu\Psi_\mu \ , \label{eqn:const_1}\\
    G_{\mu\nu}\gamma^\mu\Psi^\nu=&-3m^2\gamma^\mu\Psi_\mu+2i\(\partial_\mu m\)\gamma^{\mu\nu}\Psi_\nu \ , \label{eqn:const_2} 
\end{align}
where $G_{\mu\nu}$ is the Einstein tensor. 
Using the constraint (\ref{eqn:const_1}), the EoM (\ref{eqn:gr_eom_1}) can be rewritten as
\begin{equation}
	\cancel{\nabla}\Psi^\mu-\nabla^\mu\(\gamma^\rho\Psi_\rho\)=-\frac{im}{2}\(\gamma^{\mu\rho}\Psi_\rho+3\Psi^\mu\) \ .
\label{eqn:gr_eom_2}
\end{equation}

We now specify this general gravitino setup to the FLRW spacetime (\ref{eqn:flrw}). In this case, the starting action (\ref{eqn:L_psi}) can be canonically normalised through the field redefinition
\begin{equation}
	\Psi_\mu \equiv \frac{1}{\sqrt{a}}\psi_\mu \ ,
\label{eqn:psi_cn}
\end{equation}

\noindent and it takes the form\footnote{In formula (\ref{eqn:psi_action_cn}) the indices are contracted using the flat, Minkowski metric $\eta_{\mu\nu}$, as showed by the use of the flat Dirac matrices $\bar{\gamma}^\mu$.}
\begin{equation}
	S_\psi=\frac{1}{2}\int d^4x\[-i\bpsi_\mu\fgamma^{\mu\nu\rho}\partial_\nu\psi_\rho-am\bpsi_\mu\fgamma^{\mu\nu}\psi_\nu+2iaH\bpsi_\rho\gamma^\rho\psi_0\] \ .
\label{eqn:psi_action_cn}
\end{equation}

The EoM (\ref{eqn:gr_eom_1})-(\ref{eqn:gr_eom_2}) take the form
\begin{align}
	i\,\fgamma^{\mu\nu\rho}\partial_\nu\psi_\rho+am\fgamma^{\mu\rho}\psi_\rho+iaH\left(\eta^{\mu 0}\fgamma^\rho\psi_\rho-\fgamma^\mu\psi^0\right)=&0 \ , \label{eqn:gr_eom_flrw_1} \\
    \left(\cancel{\partial}+iam\right)\psi^\mu-\partial^\mu\left(\fgamma^\rho\psi_\rho\right)+aH\,\fgamma^\mu\psi^0+\frac{a}{2}\left[im+H\,\fgamma^0\right]\fgamma^\mu\left(\fgamma^\rho\psi_\rho\right)=&0 \ . \label{eqn:gr_eom_flrw_2}
\end{align}

\noindent Using 
\begin{align}
	R=&-6\frac{a''}{a^3} \quad , &
	G_{\mu\nu}=&a^2\left[\left(\frac{R}{3}+4H^2\right)\delta^0_\mu\delta^0_\nu-\left(\frac{R}{3}+H^2\right)\eta_{\mu\nu}\right] \ , \label{eqn:flrw_R_G}
\end{align}

\noindent and assuming the scalar field $\phi$ in eq. (\ref{eqn:L_psi}) depends only on time, as appropriate for a FLRW spacetime, the constraints (\ref{eqn:const_1})-(\ref{eqn:const_2}) become
\begin{align}
	\partial_\mu\psi^\mu=&\cancel{\partial}\left(\fgamma^\rho\psi_\rho\right)-2aH\,\psi^0-\frac{i}{2}\left(3am+iaH\,\fgamma^0\right)\left(\fgamma^\rho\psi_\rho\right) \ , \label{eqn:const_1_flrw} \\
    \fgamma^0\psi_0=& C(\tau)\,\fgamma^i\psi_i \ . \label{eqn:const_2_flrw}\
\end{align}

\noindent The function $C(\tau)$ has the expression \cite{Kolb:2021xfn}
\begin{equation}
	C(\tau)\equiv\frac{\nicefrac{R}{3}+H^2-3m^2}{3\left(H^2+m^2\right)}+i\frac{2m'}{3a\left(H^2+m^2\right)}\,\fgamma_0\equiv C_\textup{R}(\tau)+i\,C_\textup{I}(\tau)\,\fgamma_0 \ , \label{eqn:const_coeff}
\end{equation}

\noindent and is related to the sound speed parameter $c_s^2\equiv\cR^2+\cI^2$.

\subsection{The helicity eigenstates decomposition and the longitudinal gravitino}

When supersymmetry is broken, the gravitino has four degrees of freedom, corresponding to four helicity eigenstates: two of helicities $\pm \nicefrac{3}{2}$, called "transverse" gravitino, and two of helicities $\pm \nicefrac{1}{2}$, called "longitudinal" gravitino  \cite{Kallosh:1999jj, Kallosh:2000ve, Nilles:2001fg}. The longitudinal component provides the additional degrees of freedom that the massive gravitino has with respect to the massless one and it corresponds to the Goldstino fermion,  absorbed by the gravitino in the unitary gauge via the super-Higgs mechanism. As a consequence, while the transverse gravitino components are suppressed by the Planck mass, the longitudinal ones couple more strongly to matter, with a coupling that is instead suppressed by the (lower) supersymmetry breaking scale. In analogy to the case of the massive gauge field  discussed in Section \ref{section:spin_1}, the gravitino equivalence theorem \cite{Fayet:1986zc, Casalbuoni:1988kv, Casalbuoni:1988qd} establishes the matching, in the high energy limit, between the  gravitino amplitudes of the longitudinal component and the Goldstino amplitudes. 

It is necessary to properly identify the transverse and longitudinal components starting from the original Rarita--Schwinger field $\psi_\mu$. Working in momentum space via\footnote{See Appendix \ref{app:hed_algebra} for more detail about the notation and the algebraic rules.}
\begin{equation}
	\psi_\mu(\tau,\vec{x})=\int\frac{d^3k}{(2\pi)^3}\,\psi_{\mu,k}(\tau)e^{i\vec{k}\cdot\vec{x}} \ , 
\label{eqn:psi_mom_space}
\end{equation}

\noindent the space components $\psi_{i,k}$ of the gravitino can be decomposed as \cite{Kallosh:2000ve, Nilles:2001fg}
\begin{equation}
	\psi_{i,k}=\psi_{i,k}^\textup{t}+\left(P_\gamma\right)_i\,\vec{\fgamma}\cdot\vec{\psi}_k+\left(P_k\right)_i\,\vec{k}\cdot\vec{\psi}_k \ ,
\label{eqn:psi_i_dec_1}
\end{equation}

\noindent where we have introduced the two projectors
\begin{align}
	\left(P_\gamma\right)_i\equiv&\frac{1}{2}\left[\fgamma_i-\frac{k_i}{\mk}\left(\vec{k}\cdot\vec{\fgamma}\right)\right] \ , &
	\left(P_k\right)_i\equiv&-\frac{1}{2\mk}\left[3k_i+\fgamma_i\left(\vk\cdot\vfgamma\right)\right] \ ,
\label{eqn:psi_proj}
\end{align}

\noindent and $\psi^t_{i,k}$ is such that 
\begin{equation}
    \fgamma^i\psi_{i,k}^\textup{t}=k^i\psi_{i,k}^\textup{t}=0,
\end{equation} 

\noindent implying
\begin{equation}
  \left(P_\gamma\right)^i\psi_{i,k}^\textup{t}=\left(P_k\right)^i\psi_{i,k}^\textup{t}=0 \ .
\end{equation}

The time component $\psi_{0,k}$ is connected to the space ones via the constraint (\ref{eqn:const_2_flrw}), which in momentum space becomes simply
\begin{equation}
	\fgamma^0\psi_{0,k}=C\vfgamma\cdot\vpsi_k \ .
\label{eqn:const_2_k}
\end{equation}

\noindent The constraint (\ref{eqn:const_1_flrw}) takes instead the form
\begin{equation}
	\vk\cdot\vpsi_k=\left[-\vk\cdot\vfgamma+am+iaH\fgamma^0\right]\vfgamma\cdot\vpsi \ ,
\label{eqn:const_1_k}
\end{equation}

\noindent so that the decomposition (\ref{eqn:psi_i_dec_1}) can be rewritten in terms of $\psi_{i,k}^\textup{t}$ and $\vfgamma\cdot \vpsi_k\equiv\theta$ as
\begin{align}
	\psi_{i,k}=&\psi_{i,k}^\textup{t}+\hO_i\theta \ , \label{eqn:psi_dec}\\
	\hO_i\equiv&\frac{1}{\mk}\left\{k_i\left(\vk\cdot\vfgamma\right)-\frac{1}{2}\left[3k_i+\fgamma_i\left(\vk\cdot\vfgamma\right)\right]\left(am+iaH\fgamma^0\right)\right\}\theta \ . \label{eqn:psi_dec_O}
\end{align}

\noindent The field $\psi_{i,k}^\textup{t}$ corresponds to the transverse, helicity-$\pm \nicefrac{3}{2}$ components of the gravitino, the field $\theta$ to the longitudinal, helicities-$\pm \nicefrac{1}{2}$ one and the decomposition (\ref{eqn:psi_dec}) allows to project on them any function of the gravitino.

\subsection{The hamiltonian and canonical quantisation}

One now performs the Fock space analysis we discussed in the previous sections for the hamiltonians of the transverse and longitudinal gravitinos. Starting from the canonically normalised action (\ref{eqn:psi_action_cn}) and applying the decomposition (\ref{eqn:psi_dec})-(\ref{eqn:psi_dec_O}), one finds, in agreement with \cite{Nilles:2001fg, Kallosh:1999jj, Kallosh:2000ve},
\begin{align}
	\mathcal{L}_\psi=&\mathcal{L}_\textup{t}+\mathcal{L}_\theta \quad , \quad {\rm where} \\
	\mathcal{L}_\textup{t}=&\frac{1}{2}\sum_{i=1}^3\bpsi_{i,k}^\textup{t}\left[i\fgamma^0\partial_0-\kg-am\right]\psi_{i,k}^\textup{t} \ ,\\
	\mathcal{L}_\theta=&\frac{\Upsilon a^2}{4\mk}\[i\btheta\fgamma^0\partial_0\theta-\btheta\kg C\theta +\frac{am}{2}\btheta\theta+\frac{3}{2}\(amC_\textup{R}+aHC_\textup{I}\)\btheta\theta\] \ ,
\label{eqn:L_theta_0}
\end{align}

\noindent where the coefficient $\Upsilon$ is
\begin{equation}
    \Upsilon \equiv 3\(m^2+H^2\) \ .
\end{equation}

Therefore, the dynamics of the transverse and longitudinal components decouple from each other. The action of the former takes exactly the form of a spin-$\nicefrac{1}{2}$ fermion in FLRW. Also the action for the latter has that form, but the time-dependent mass parameter and the sound speed coefficient are non-trivial. The lagrangian (\ref{eqn:L_theta_0}) can be canonically normalised via\footnote{The $\fgamma^0$ matrix that appears in (\ref{eqn:theta_cn}) is needed to canonically normalise the spatial kinetic term, so that in the Minkowski limit, in which $C\to-1$, we recover the lagrangian of a standard spin-$\nicefrac{1}{2}$ fermion.}
\begin{equation}
    \theta=\sqrt{\frac{\Upsilon a^2}{2\vk^2}}\fgamma^0\vartheta \ ,
\label{eqn:theta_cn}
\end{equation}

\noindent and it becomes
\begin{equation}
    \mathcal{L}_\vartheta=\frac{1}{2}\[i\bar{\vartheta}\fgamma^0\partial_0\vartheta+\bar{\vartheta}\kg C\vartheta-aM\bar{\vartheta}\vartheta\] \ ,
\label{eqn:L_theta}
\end{equation}

\noindent where we defined the effective mass parameter
\begin{equation}
    M \equiv -\[\frac{m}{2}+\frac{3}{2}\(mC_\textup{R}+HC_\textup{I}\)\] \ .
\end{equation}

The hamiltonians for the transverse and longitudinal components are found to be 
\begin{align}
	H_\psi=&\int d^3k\[\mathcal{H}_\textup{t}+\mathcal{H}_\vartheta\] \ ,\\
	\mathcal{H}_\textup{t}=&\frac{1}{2}\sum_{i=1}^3\bpsi_{i,k}^\textup{t}\[\kg+am\]\psi_{i,k}^\textup{t} \ ,\\
	\mathcal{H}_\vartheta=&\frac{1}{2}\[-\bar{\vartheta}\kg C\vartheta +aM\bar{\vartheta}\vartheta\] \ . 
\label{eqn:H_theta}
\end{align}

\noindent Starting from (\ref{eqn:H_theta}), one can perform the quantisation of the system, following \cite{Nilles:2001fg, Peloso:2000hy, Chung:2011ck}. Using the conjugate momentum of $\vartheta$,
\begin{equation}
    \Pi_\vartheta \equiv \frac{\partial\L_\vartheta}{\partial\partial_0\vartheta}=\frac{i}{2}\vartheta^\dagger,
\end{equation}

\noindent we impose the canonical commutation relations 
\begin{equation}
\begin{aligned}
	\left\{\vartheta(\tau,\vx),\Pi_\vartheta(\tau,\vec{y})\right\}=&\delta^{(3)}\(\vx-\vec{y}\) \ , \\
 \left\{\vartheta(\tau,\vx),\vartheta(\tau,\vec{y})\right\}=&\left\{\Pi_\vartheta(\tau,\vx),\Pi_\vartheta(\tau,\vec{y})\right\}=0 \ .
\end{aligned}
\label{eqn:psi_comm_rel_ps}
\end{equation}

\noindent These relations allow to define a decomposition of $\vartheta$ in momentum space in terms of ladder operators as
\begin{equation}
	\vartheta(\tau,\vec{x})=\int \frac{d^3k}{(2\pi)^{\nicefrac{3}{2}}}\sum_r e^{i\vk\cdot\vx}\[\vartheta_{r,\bm{k}}a_{r,\bm{k}}+v_{r,\bm{k}}a_{r,-\bm{k}}^\dagger\] \ ,
\label{eqn:theta_ladder_dec}
\end{equation}

\noindent where the index $r$ spans the two possible polarisations $\pm \nicefrac{1}{2}$. The two mode functions $\vartheta_{r,\bm{k}}$ and $v_{r,\bm{k}}$ are normalised as
\begin{equation}
	\vartheta_{r,\bm{k}}^\dagger\vartheta_{s,\bm{k}}=\delta_{rs}=v_{r,\bm{k}}^\dagger v_{s,\bm{k}} \ ,
\label{eqn:theta_mf_norm}
\end{equation}
\noindent and are such that
\begin{equation}
	\bar{v}_{r,\bm{k}}=\vartheta_{r,-\bm{k}}^\textup{T}\mathcal{C} \ , 
\label{eqn:maj_cond_mod_f}
\end{equation}

\noindent where $\mathcal{C}$ is the charge conjugation matrix.
Working in the reference frame in which $\vk=\(0,0,k\)$, we can decompose the mode function $\vartheta_{r,\bm{k}}$ in two-component notation as \cite{Nilles:2001fg}
\begin{equation}
	\vartheta_{r,\bm{k}}=\begin{pmatrix}\vartheta_A(r,\bm{k})\\\vartheta_B(r,\bm{k})\end{pmatrix}=\frac{1}{2}\begin{pmatrix}\(\vartheta_+-r\vartheta_-\)\psi_r\\ \(\vartheta_+ + r\vartheta_-\)\psi_r\end{pmatrix} \ ,
\label{eqn:theta_dec}
\end{equation}

\noindent where the $\vk$-dependence in $\vartheta_{\pm}$ is left implicit. The two-component vectors $\psi_r\equiv (\psi_+,\psi_-)$ are taken to be the eigenvectors of the helicity operator $\vec{\sigma}\cdot\hat{k}=\sigma^3$, namely
\begin{equation}
	\psi_+=\begin{pmatrix}1\\0\\\end{pmatrix} \ , \quad \psi_-=\begin{pmatrix}0\\1\\\end{pmatrix} \ , \qquad \qquad \sigma^3\psi_r=r\psi_r \ .
\label{eqn:psi_r_def}
\end{equation}

\noindent Imposing canonical commutation relations on the ladder operators introduced in (\ref{eqn:theta_ladder_dec}) 
\begin{equation}
	\left\{a_{r,\bm{k}},a_{s,\bm{q}}^\dagger\right\}=\delta_{rs}\delta^{(3)}\(\vk-\vec{q}\) \ , \qquad \qquad \left\{a_{r,\bm{k}},a_{s,\bm{q}}\right\}=0=\left\{a_{r,\bm{k}}^\dagger,a_{s,\bm{q}}^\dagger\right\} \ ,
\label{eqn:a_ad_cr}
\end{equation}

\noindent one finds that the mode functions $\vartheta_\pm$ have to be normalised such that
\begin{equation}
	\abs{\vartheta_+}^2+\abs{\vartheta_-}^2=2 \ .
\label{eqn:theta_pm_norm}
\end{equation}

\noindent From the lagrangian (\ref{eqn:L_theta}) one finds that $\vartheta_\pm$ satisfy the following EoM\footnote{To go from four to two components we use the following Weyl representation of the Dirac matrices:
\begin{equation}
    \fgamma^{\mu}=\begin{pmatrix} 0 & \sigma^\mu \\ \bar{\sigma}^\mu & 0\end{pmatrix},
\label{eqn:4to2_gamma}
\end{equation}
\noindent where $\sigma^{\mu}=\(\mathbb{1},\vec{\sigma}\)$ and $\bar{\sigma}^{\mu}=\(\mathbb{1},-\vec{\sigma}\)$.}: 
\begin{equation}
	{\vartheta}'_\pm = -i\(C_\textup{R}\mp i C_\textup{I}\)k \vartheta_{\mp}\mp i aM \vartheta_\pm \ .
\label{eqn:theta_pm_eom}
\end{equation}

In this notation, the hamiltonian (\ref{eqn:H_theta}) becomes
\begin{equation}
	\H_\vartheta=\frac{1}{2}\sum_r\[\begin{pmatrix} a_{r,\bm{k}}^\dagger&a_{-r,-\bm{k}}\end{pmatrix}\begin{pmatrix}\mathcal{E}_k & \mathcal{F}_k^\dagger \\ \mathcal{F}_k&-\mathcal{E}_k\end{pmatrix}\begin{pmatrix} a_{r,\bm{k}}\\a_{-r,-\bm{k}}^\dagger\end{pmatrix}\] \ , 
\label{eqn:H_theta_ladder}
\end{equation}

\noindent where
\begin{equation}
	\begin{aligned}
		\mathcal{E}_k=&-\frac{\cR k}{2}\(\vartheta_+^*\vartheta_-+\vartheta_+\vartheta_-^*\)+\frac{i}{2}\cI k \(\vartheta_+^*\vartheta_- - \vartheta_+\vartheta_-^*\)+\frac{aM}{2}\(\abs{\vartheta_+}^2-\abs{\vartheta_-}^2\) \ , \\
		\mathcal{F}_k=&-\frac{\cR k}{2}\(\vartheta_+^2-\vartheta_-^2\)-\frac{i}{2}\cI k \(\vartheta_+^2+\vartheta_-^2\)-aM\vartheta_+\vartheta_- \ .
	\end{aligned}
\label{eqn:theta_E_F}
\end{equation}

Following the procedure of the previous sections, one now applies a Bogolyubov transformation. In this case, the Bogolyubov decomposition is performed on the mode functions $\vartheta_\pm$ according to  \cite{Nilles:2001fg, Peloso:2000hy, Chung:2011ck}
\begin{equation}
\begin{aligned}
	\vartheta_+ \ = \ &c_+(k) \ \alpha_{\bm{k}}-c_-(k) \ \beta_{\bm{k}} \ , \\
	\vartheta_- \ = \ &c_-(k) \ \alpha_{\bm{k}}+c_+(k) \ \beta_{\bm{k}} \ .
\end{aligned}
\label{eqn:theta_bog_tr}
\end{equation}

\noindent As a consequence of the corresponding EoM, the Bogolyubov coefficients $\alpha_{\bm{k}}$ and $\beta_{\bm{k}}$ are respectively even and odd in $\vk$:
\begin{align}
    \alpha_{-\bm{k}}=\alpha_{\bm{k}} \ , &&  \beta_{-\bm{k}}=-\beta_{\bm{k}} \ ,
\label{eqn:theta_bog_parity}
\end{align}
\noindent while the coefficients $c_\pm$ are real, normalised as
\begin{equation}
	c^2_+({k})+c^2_-({k})=2 \ .
\label{eqn:c_pm_norm}
\end{equation}

\noindent They also have the parity properties
\begin{align}
    c_+(-k)=c_+(k) \ , && c_-(-k)=-c_-(k) \ .
\label{eqn:cpm_parity}
\end{align}

\noindent These are the minimal requirements for (\ref{eqn:theta_bog_tr}) to be a well-defined Bogolyubov decomposition, with the coefficients  normalised as
\begin{equation}
	\abs{\alpha_{\bm{k}}}^2+\abs{\beta_{\bm{k}}}^2=1 \ ,
\label{eqn:bog_coeff_norm}
\end{equation}

\noindent in order for the commutation relations (\ref{eqn:a_ad_cr}) to be preserved by (\ref{eqn:theta_bog_tr}). 

The instantaneous hamiltonian (\ref{eqn:H_theta_ladder}) becomes
\begin{equation}
    \H_\vartheta=\frac{1}{2}\sum_r\begin{pmatrix} \ta_{r,\bm{k}}^\dagger&\ta_{-r,-\bm{k}}\end{pmatrix}\begin{pmatrix}A_k & B_k\\ B_k^\dagger &-A_k\end{pmatrix}\begin{pmatrix} \ta_{r,\bm{k}}\\ \ta^\dagger_{-r,-\bm{k}},\end{pmatrix} \ ,
\label{eqn:H_theta_1}
\end{equation}

\noindent where the new set of ladder operators is defined as
\begin{equation}
	\begin{pmatrix} \ta_{r,\bm{k}}\\ \ta^\dagger_{-r,-\bm{k}},\end{pmatrix}= \begin{pmatrix}\alpha_{\bm{k}}& -\beta_{\bm{k}}^\dagger\\\beta_{\bm{k}}&\alpha_{\bm{k}}^\dagger\end{pmatrix}\begin{pmatrix} a_{r,\bm{k}}\\a_{-r,-\bm{k}}^\dagger\end{pmatrix} \ ,
\label{eqn:theta_bog_tr_1}
\end{equation}

\noindent while the functions $A_k$ and $B_k$ are
\begin{equation}
\begin{aligned}
	A_k=&-\[c_+(k)c_-(k)\]\cR k+\[\frac{c_+^2(k)-c_-^2(k)}{2}\] aM \ ,\\
	B_k=&-\[\frac{c_+^2(k)-c_-^2(k)}{2}\]\cR k+i\cI k-\[c_+(k)c_-(k)\] aM \ .
\end{aligned}
\label{eqn:theta_AB}
\end{equation}

Note that, contrary to the previous cases of lower spins, the bare Bogolyubov decomposition (\ref{eqn:theta_bog_tr}) is not enough to diagonalize the instantaneous hamiltonian of the longitudinal gravitino. It is in fact not possible to solve for $B_k=0$ in terms of the coefficients $c_\pm$ because of the non-vanishing imaginary part\footnote{One possibility to avoid such difference with the previous cases would be to reabsorb the complex phase of the sound speed into a field redefinition of $\vartheta$. In this way the sound speed parameter becomes real and there exist a choice of $c_\pm$ in (\ref{eqn:theta_bog_tr}) that actually diagonalizes the hamiltonian right away. However, such choice does not change any of the final results of this analysis. \label{fn:theta_ham}}. In order to achieve the desired diagonal form, one needs a further Bogolyubov transformation, defined by
\begin{equation}
    \begin{pmatrix} \hat{b}_{r,\bm{k}}\\ \hat{b}^\dagger_{-r,-\bm{k}}\end{pmatrix}= \begin{pmatrix}\halpha_{\bm{k}}& -\hbeta_{\bm{k}}^\dagger\\ \hbeta_{\bm{k}}&\halpha_{\bm{k}}^\dagger\end{pmatrix}\begin{pmatrix} \ta_{r,\bm{k}}\\ \ta_{-r,-\bm{k}}^\dagger\end{pmatrix} \ , 
\label{eqn:theta_bog_tr_2}
\end{equation}

\noindent with\footnote{In (\ref{eqn:H_hbog}) we write only the modulus of the Bogolyubov coefficients because the equation defining them can only be solved up to a phase which is however unphysical.} 
\begin{align}
     \halpha_{\bm{k}}^\dagger=&-\frac{B_k}{\omega-A_k}\hbeta_{\bm{k}}, & \abs{\halpha_{\bm{k}}}^2=&\frac{\omega+A_k}{2\omega},& \abs{\hbeta_{\bm{k}}}^2=\frac{\omega-A_k}{2\omega} \ ,
\label{eqn:H_hbog}
\end{align}

\noindent where $\omega$ is the frequency defined by
\begin{align}
    \omega^2=&c_s^2k^2+a^2M^2 \ , & c_s^2=\cR^2+\cI^2 \ 
\end{align}
and $c_s$ is the longitudinal gravitino sound speed. 
\noindent The resulting diagonal hamiltonian is then
\begin{equation}
    \H_\vartheta=\frac{\omega}{2}\sum_r\(\hat{b}_{r,\bm{k}}^\dagger \hat{b}_{r,\bm{k}}-\hat{b}_{r,\bm{k}} \hat{b}_{r,\bm{k}}^\dagger\) \ .
\label{eqn:H_theta_diag}
\end{equation}

\noindent Note that this result is independent of the specific choice of the coefficients $c_\pm$ in the original Bogolyubov transformation (\ref{eqn:theta_bog_tr}) (it only depends on the combination fixed by the normalisation (\ref{eqn:c_pm_norm})).

\subsection{The energy-momentum tensor}

We now compare the instantaneous hamiltonian (\ref{eqn:H_theta_diag})  with the actual physical energy of the longitudinal gravitino, encoded in its energy-momentum tensor. To do this, we start from the gravitino sector of lagrangian (\ref{eqn:L_psi}) and employ the definition (\ref{eqn:emt_def_e}). On-shell one finds 
\begin{equation}
\begin{aligned}
	T^{\mu\nu}=&-\frac{i}{2}\bPsi_\rho\gamma^{\rho\alpha(\mu}\({\nabla}_\alpha\Psi^{\nu)}-{\nabla}^{\nu)}\Psi_\alpha\)-\frac{m}{2}\bPsi_\rho\gamma^{\rho(\mu}\Psi^{\nu)}\\
		&+\frac{i}{2}\nabla_\rho\(\bPsi^\rho\gamma^{(\mu}\Psi^{\nu)}\)+\frac{i}{2}\nabla^{(\mu}\left(\bPsi^{\nu)}\gamma^\rho\Psi_\rho\)+\frac{i}{2}g^{\mu\nu}\,\nabla_\rho\left(\bPsi_\alpha\gamma^\alpha\Psi^\rho\right) \ .
\end{aligned}
\label{eqn:emt_full}
\end{equation}

\noindent 
In Appendix~\ref{app:emt} we explicitly show that this tensor is, on-shell, covariantly conserved. From  (\ref{eqn:emt_full}) one can compute the energy operator of the gravitino in an FLRW spacetime. In terms of the canonically normalised field $\psi_\mu$ defined in (\ref{eqn:psi_cn}), this reads
\begin{equation}
    E_\psi=\int d^3x \g \ T^0{}_0=\frac{1}{2}\int d^3x\left\{i\bpsi_i\fgamma^{i j l}{\partial}_j\psi_l+am\bpsi_i\fgamma^{ij}\psi_j\right\} \ .
\label{eqn:gr_eom_flrw_0_comp}
\end{equation}

\noindent One can then decompose this energy in terms of the transverse and longitudinal components by going to momentum space via (\ref{eqn:psi_mom_space}) and apply the decomposition (\ref{eqn:const_2_k})-(\ref{eqn:psi_dec}). The resulting energy operator is
\begin{align}
    E_\psi=&\int d^3k\[\E_\textup{t}+\E_\theta\] , \quad {\rm where } \  \label{eqn:psi_E_flrw_tot}\\
    \E_\textup{t}=&\frac{1}{2}\sum_{j=1}^3\[\bpsi_{j,k}^\textup{t}\kg\psi_{j,k}^\textup{t}+am\bpsi_{j,k}^\textup{t}\psi_{j,k}^\textup{t}\] \ ,  \label{eqn:psi_ET}\\
	\E_\vartheta=&\frac{\Upsilon a^2}{4\vk^2}\btheta\[-\kg\(\frac{m-\frac{i}{3}H\fgamma^0}{m-iH\fgamma^0}\)+am\]\theta \ . \label{eqn:E_theta_0}
\end{align}

\noindent We observe that the longitudinal gravitino energy density (\ref{eqn:E_theta_0}) agrees with \cite{Kallosh:1999jj}. Performing the redefinition (\ref{eqn:theta_cn}) and introducing the quantity 
\begin{equation}
	D=d_\textup{R}+id_\textup{I}\fgamma^0\equiv\frac{3m^2+H^2}{3\(m^2+H^2\)}+i\frac{2mH}{3\(m^2+H^2\)}\fgamma^0 \ ,
\label{eqn:theta_D}
\end{equation}

\noindent it becomes 
\begin{equation}
	\E_\vartheta=\frac{1}{2}\bar{\vartheta}\[\kg D+am\]\vartheta \ .
\label{eqn:E_theta_D}
\end{equation}

We see therefore that, while the transverse gravitino behaves precisely as the spin-$\nicefrac{1}{2}$ fermion discussed in Section \ref{section:spin_12}, the energy operator (\ref{eqn:E_theta_D}) and the instantaneous hamiltonian (\ref{eqn:H_theta}) of the longitudinal component do not coincide. They have though the same structure, which allows to easily perform the quantisation and the diagonalisation. The final, diagonal energy density operator for the longitudinal gravitino is given by
\begin{equation}
	\E_\vartheta=\frac{\omega_d}{2}\sum_r\(\hat{a}_{r,\bm{k}}^\dagger \hat{a}_{r,\bm{k}}-\hat{a}_{r,\bm{k}} \hat{a}_{r,\bm{k}}^\dagger\) \ ,
\label{eqn:E_theta_diag}
\end{equation}

\noindent where the diagonal frequency is given by
\begin{align}
	\omega_d^2=d_s^2k^2+a^2m^2 \ , \quad {\rm where} && d_s^2=d_R^2+d_I^2=\frac{m^2+\nicefrac{H^2}{9}}{m^2+H^2} \ . 
\label{eqn:theta_wd}
\end{align}

\noindent The diagonal ladder operators are obtained in terms of the original ones in (\ref{eqn:theta_ladder_dec}) as
\begin{equation}
    \begin{pmatrix} \hat{a}_{r,\bm{k}}\\ \hat{a}^\dagger_{-r,-\bm{k}},\end{pmatrix}=\begin{pmatrix}x_{\bm{k}}& -y_{\bm{k}}^\dagger\\ y_{\bm{k}}&x_{\bm{k}}^\dagger\end{pmatrix}\begin{pmatrix} \ta_{r,\bm{k}}\\ \ta_{-r,-\bm{k}}^\dagger\end{pmatrix} =\begin{pmatrix}x_{\bm{k}}& -y_{\bm{k}}^\dagger\\ y_{\bm{k}}&x_{\bm{k}}^\dagger\end{pmatrix}\begin{pmatrix}\alpha_{\bm{k}}& -\beta_{\bm{k}}^\dagger\\\beta_{\bm{k}}&\alpha_{\bm{k}}^\dagger\end{pmatrix}\begin{pmatrix} a_{r,\bm{k}}\\a_{-r,-\bm{k}}^\dagger\end{pmatrix} \ ,
\label{eqn:E_diag_lad}
\end{equation}

\noindent where the Bogolyubov coefficients $x_k$ and $y_k$ - the energy operator counterpart of the $\halpha_k$ and $\hbeta_k$ introduced for the instantaneous hamiltonian in (\ref{eqn:theta_bog_tr_2})-(\ref{eqn:H_hbog}) - are defined by\footnote{As for (\ref{eqn:H_hbog}), the Bogolyubov coefficients are defined up to an unphysical  phase.}
\begin{align}
    x_{\bm{k}}^\dagger=&-\frac{B_k^\textup{D}}{\omega_d-A_k^\textup{D}}y_{\bm{k}} \ , & \abs{x_{\bm{k}}}^2=&\frac{\omega_d+A_k^\textup{D}}{2\omega_d} \ ,& \abs{y_{\bm{k}}}^2=&\frac{\omega_d-A_k^\textup{D}}{2\omega_d} \ , 
\label{eqn:E_hbog}
\end{align}

\noindent with 
\begin{align}
    A_k^\textup{D} = &[c_+(k)c_-(k)]\dr k+\(\frac{c^2_+(k)-c^2_-(k)}{2}\)am \ , \\
    B_k^\textup{D} = &\(\frac{c^2_+(k)-c^2_-(k)}{2}\)\dr k-[c_+(k)c_-(k)]am -i\di k \ .
\end{align}

\subsection{Discussion on the two Fock spaces and a puzzle}

We have now all the ingredients to quantify also for the longitudinal gravitino the difference between the two diagonal Fock spaces of the instantaneous hamiltonian and the energy operator. The Bogolyubov transformation between the two is obtained combining equations (\ref{eqn:theta_bog_tr_2}) and (\ref{eqn:E_diag_lad}), and it reads
\begin{equation}
    \begin{pmatrix} \hat{a}_{r,\bm{k}}\\ \hat{a}_{-r,-\bm{k}},^\dagger\end{pmatrix}=\begin{pmatrix}\mathscr{A}_{\bm{k}}& -\mathscr{B}_{\bm{k}}^\dagger\\ \mathscr{B}_{\bm{k}}&\mathscr{A}_{\bm{k}}^\dagger\end{pmatrix}\begin{pmatrix} \hat{b}_{r,\bm{k}}\\ \hat{b}_{-r,-\bm{k}},^\dagger\end{pmatrix} \ , 
\label{eqn:psi_comp_fs}
\end{equation}

\noindent with
\begin{align}
    \mathscr{A}_{\bm{k}}=x_{\bm{k}}\halpha_{\bm{k}}^\dagger+y_{\bm{k}}^\dagger\hbeta_{\bm{k}} \ , && \mathscr{B}_{\bm{k}}=y_{\bm{k}}\halpha_{\bm{k}}^\dagger-x_{\bm{k}}^\dagger\hbeta_{\bm{k}} \ .
\label{eqn:psi_fs_bog}
\end{align}

Combining equations (\ref{eqn:H_hbog}) and (\ref{eqn:E_hbog}), one obtains
\begin{align}
    \abs{\mathscr{A}_{\bm{k}}}^2=&\frac{1}{2}\[1-\frac{\(\cR\dr+\cI\di\)}{\omega\omega_d}k^2+\frac{a^2mM}{\omega\omega_d}\],\\
    \abs{\mathscr{B}_{\bm{k}}}^2=&\frac{1}{2}\[1+\frac{\(\cR\dr+\cI\di\)}{\omega\omega_d}k^2-\frac{a^2mM}{\omega\omega_d}\] \ . \label{eq:calB2}
\end{align}

\noindent Note that, once again, these formulas are independent of the particular choice of coefficients $c_\pm$ in (\ref{eqn:theta_bog_tr}). 
We can then study the high-momentum UV limit of this transformation.  Contrary to the previous cases of lower spin fields, this time the two Fock spaces are different even in the UV, since 
\begin{equation}
    \lim_{k\to\infty}\abs{\mathscr{B}_{\bm{k}}}^2=\frac{1}{2}\(1+\frac{\cR\dr+\cI\di}{c_sd_s}\) \ . \label{eq:calB3} 
\end{equation}

\noindent The term proportional to the sound speed parameters is time-dependent such that the transformation in eq. (\ref{eqn:psi_comp_fs}) is non trivial, even for large momenta. Since in the high-momentum limit one expects, intuitively, the effects of spacetime dynamics to be small on particles of any spin, we do not understand this difference with respect to particles of the other spins considered previously.  

\section{Comments on the initial state and the zero-point energy}\label{section:norm_ord}

Let us now comment about the natural choice for the initial state and corresponding zero-point energy for the instantaneous hamiltonian and the energy operator. For definiteness, and since this is the most delicate example, we consider the case of the spin $3/2$ Rarita--Schwinger field. Essentially all our discussion applies also with appropriate changes for the other spins.

The diagonal Fock space of the instantaneous hamiltonian $\H_\vartheta$ (\ref{eqn:H_theta_diag}) is the one usually used in order to define the number of particles, whereas the Fock space of the energy operator $\E_\vartheta$ (\ref{eqn:E_theta_diag}) defines the energy. The two associated operators are, respectively,
\begin{align}
 N_\vartheta =\sum_r  \hat{b}_{r,{\bm{k}}}^\dagger \hat{b}_{r,{\bm{k}}}, && \E_\vartheta  =     \omega_d \  \sum_r  \(\hat{a}_{r,{\bm{k}}}^\dagger \hat{a}_{r,{\bm{k}}} - \frac{1}{2} \) \ . \label{eq:comm1}
\end{align}

\noindent As standard in cosmology, we work in the Heisenberg picture, where states are time-independent whereas the operators evolve in time. Notice that the (hatted) operators diagonalizing the two Fock spaces in discussion are time-dependent and so are therefore the operators in (\ref{eq:comm1}). For this reason we have to specify at what time the state (\ref{eq:comm2}) is evaluated. We choose to define it at some initial time $\tau_0$ in which the mode is deeply inside the horizon.

Thus, according to (\ref{eq:comm1}), a state of zero particles is defined, at initial time $\tau_0$, by  
\begin{equation}
\hat{b}_{r,{\bm{k}}} (\tau_0) \ | \chi_0 \rangle \ = 0 \ . \label{eq:comm2}   
\end{equation}

\noindent Since on this state the instantaneous gravitino hamiltonian $\H_\vartheta$ (\ref{eqn:H_theta_diag}) is minimized, the state $\ket{\chi_0}$ is the Bunch--Davies vacuum, is defined in (\ref{eqn:BD}).

We can then compute the energy of this state again from (\ref{eq:comm1}), obtaining
\begin{equation}
\langle  \chi_0 | \E_\vartheta (\tau_0) | \chi_0 \rangle =
\omega_d \ \langle \chi_0 | \sum_r  \(\hat{a}_{r,{\bm{k}}}^\dagger \hat{a}_{r,{\bm{k}}} - \frac{1}{2} \)  | \chi_0 \rangle = 
\omega_d \ \(-1+ 2\abs{\mathscr{B}_{\bm{k}} (\tau_0)}^2\) V_3 \ , \label{eq:comm3} 
\end{equation}

\noindent where we used eq. (\ref{eqn:psi_comp_fs}) and defined as previously the space volume as $V_3 = \delta^3 ({\bm k}=0)$. The term with $-1$ is the zero-point energy, on which we will comment later on, whereas the term in $\abs{\mathscr{B}_{\bm{k}}}^2$ is an additional positive contribution.  The state with no particles has therefore a non-trivial energy. This also means that, defined in this manner the BD vacuum itself is not generally\footnote{The spin-$\nicefrac{1}{2}$ field in FLRW is an exception, since in this case the energy operator and the instantaneous hamiltonian do coincide, as discussed in Section \ref{section:spin_12}.} the state with minimum energy. It is instead the state which contains zero particles, according to the standard definition of number operator given in (\ref{eq:comm1}). However, this notion in curved spacetime is ambiguous and the fact that the state with no particles actually contains energy is additional evidence in this respect. Moreover, whereas for the case of spin  $0,1/2$ and $1$  states with high-momenta do not change the energy of the ground state, this is not true for spin $3/2$, as can be seen by using eq. (\ref{eq:calB2}). 

From the viewpoint of discussing backreaction of produced particles, it is more natural to start with a state of minimum energy, defined by
\begin{equation}
\hat{a}_{r,{\bm{k}}} (\tau_0) \ | \psi_0 \rangle \ = 0 \ . \label{eq:comm4} 
\end{equation}

\noindent The number of particles of this state is then computed to be
\begin{equation}
\langle  \psi_0 | N_\vartheta | \psi_0 \rangle \ =
 \ \langle \psi_0 | \sum_r  \(\hat{b}_{r,{\bm{k}}}^\dagger \hat{b}_{r,{\bm{k}}}\) | \psi_0 \rangle = 
2 \ \abs{\mathscr{B}_{\bm{k}}(\tau_0)}^2 V_3 \ . \label{eq:comm5} 
\end{equation}

\noindent The natural state with minimum energy and consequently with no backreaction (after subtracting the zero-point energy) on the geometry has therefore a non-vanishing number of ``particles''. Of course, this reflects to a large extent the known fact that the number of particles in time-dependent backgrounds is ambiguous. 
The novel puzzling fact is that, whereas for the case of spin $0,1/2$  and $1$, states with asymptotically high-momenta do not contribute to this ``number of particles'', this is not true for spin $3/2$, as seen from eq. (\ref{eq:calB3}). In order to find zero particles the sound speed condition $c_s=1$ is not enough. Indeed, one also need $d_s=1$, implying $H=0$. Only flat space seems to satisfy such requirement.  

Our discussion on the number of particles is a slightly academic one, since the total number of particles discussed above, in analogy with the case of the scalar and spin $1$ fields, is actually divergent in the UV, after integrating over all momenta. It is therefore necessary to discuss the energy instead of the number of particles.  

Until now, we have avoided discussing the zero-point energies of the different energy operators and their interpretation. Normal ordering prescription of the vacuum energy leads to divergences that should be removed by corresponding counterterms in the lagrangian. This is not completely arbitrary, since the counterterms should be operators invariant under all symmetries of the theory, most notably diffeomorphisms. Vacuum energy is related to the energy-momentum tensor via
\begin{equation}
\rho \ = \ T^{0}{}_0 \ = \ \frac{1}{a^2} T_{00}   \ . \label{eq:comm6} 
\end{equation}
If one uses for the UV cutoff a Pauli-Villars prescription with particles of mass $\Lambda$, it is easy to see that the cutoff on the comoving momentum is $ a \Lambda$.  Then a constant term $\delta \rho \sim \Lambda^4$ in the vacuum energy corresponds to a renormalization of the cosmological constant, whereas a term of the type
$\delta \rho \sim H^2 \Lambda^2$ corresponds to a renormalization of the Newton constant. This can be seen from (\ref{eqn:flrw_R_G}) , which implies $G^0{}_0 = 3 H^2$. Indeed, through Einstein eqs., $G^0{}_0$ is the contribution to the vacuum energy coming from a Einstein-Hilbert term in the action. 
The higher order terms in the derivatives of the scale factor correspond to $R^2$ and $R_{\mu \nu}^2$ terms counterterms in the effective action, that will not be needed in our following discussion. 

Normal ordering in the diagonal Fock space of the instantaneous hamiltonian $\H_\vartheta$ (\ref{eqn:H_theta_diag}) leads to a zero-point energy $-\omega V_3$, whereas a normal ordering in the diagonal Fock space of the energy operator $\E_\vartheta$ (\ref{eqn:E_theta_diag}) leads to a zero-point energy $-\omega_d V_3$. One can use the high-momentum expansion of the two frequencies 
\begin{equation}
\omega = c_s k \[1 + \mathcal{O}\(\frac{M^2}{k^2}\)\] \quad , \quad  \omega_d = d_s k \[1 + \mathcal{O}\(\frac{a^2 m^2}{k^2}\)\]  \ . \label{eq:comm7} 
\end{equation} 

\noindent The leading terms in the expansion eq. (\ref{eq:comm7}) corresponds to quartic UV divergences in the vacuum energy, whereas the first subleading term leads to quadratic divergences. We discuss here for simplicity only the leading terms. The expansion of $d_s$ in derivatives of the scale factor, equivalent to the derivative expansion in the 
effective action, is simple:
\begin{equation}
d_s = 1 - \frac{4 H^2}{9 m^2} + \cdots    \ . \label{eq:comm8}    
\end{equation}

\noindent Plugging the result back into eq. (\ref{eq:comm7}) one finds two counterterms, one proportional to $\Lambda^4$, renormalizing the cosmological constant, and a second one proportional to $\frac{H^2}{m^2}\Lambda^4$, renormalizing the Planck mass.  

On the other hand, if one would attempt to naively use the zero-point energy defined from the Fock space of $\H_\vartheta$ in the energy operator, one should use the derivative expansion of $c_s$. This is more complicated and it deals, in particular, with the time-derivatives of the gravitino mass, see eq. (\ref{eqn:const_coeff}). We were unable to identify diffeomorphism-invariant counterterms in the effective action from its expansion: we therefore conclude that the normal ordering should be performed with the Fock space of the energy operator $\E_\vartheta$ and {\it not} with the one of the instantaneous hamiltonian $\H_\vartheta$.  

To summarize our current situation, whereas the Fock space for the energy operator $T^0{}_0$ and the instantaneous hamiltonian of the canonically normalized field coincide only for the spin $1/2$ field, for the spins $0$ and $1$ the two Fock spaces match (only) in the limit of high momenta. In these cases, one can use the definition of one-particle states from the respective instantaneous hamiltonians for high-momenta. Since intuitively high-momenta are not affected by the spacetime curvature, one would have expected the result to extend also to the spin $3/2$ case. The failure of this reasoning is the main outcoming puzzle of our work. 

The conclusion of our analysis is that the BD vacuum defined as the state minimizing the instantaneous hamiltonian is not a the state of minimum energy. The minimum energy is instead the ground state of the  energy operator $T^0{}_0$, as we argued throughout this paper. This result applies to cases of fields of all spin. For example, for the case of the scalar field discussed in Section \ref{section:spin_0}, we  found that the BD vacuum has the energy (\ref{eq:energyBD}).  
Like for the case of spin $3/2$, it seems that this energy cannot be renormalized to zero by a consistent counter-term. We propose that minimizing the energy operator, eq. (\ref{eq:comm4}) in the case of spin $3/2$ and the similar ground states of the energy operator for the other spins,  should be the physical condition to impose to define a vacuum state. 
Indeed, the vacuum state should be defined to have vanishing energy at initial time after renormalization
\begin{equation}
    \langle T^0{}_0\rangle(\tau_0)=0 \ , 
\label{eqn:vev_bd_T00}
\end{equation}

\noindent even if it contains a non-vanishing number of particles in the usual terminology, as seen for the gravitino in equation (\ref{eq:comm5}). Indeed, the energy-momentum tensor is a physical observable which should be used to describe backreaction of produced particles on the spacetime geometry.

In order to perform such computations (in particular for the gravitino), it is of course necessary to perform a full renormalization of the vacuum energy. The zero-point energy coming from the normal ordering, that we commented above, is just the first term in an adiabatic expansion. We leave a detailed discussion of the renormalization and the backreaction of gravitino production on the geometry for future work.~\footnote{For thermal and other mechanisms of gravitino production, see for instance \cite{Giudice:1999yt}-\cite{Nilles:2001ry}, \cite{Bolz:2000fu},\cite{Benakli:2017whb}.}


\section{The gravitino equivalence theorem and a second puzzle}\label{section:eq_th}

As for the massive spin-1 field, an equivalence theorem holds also for the massive gravitino \cite{Fayet:1986zc, Casalbuoni:1988kv, Casalbuoni:1988qd}, stating that the longitudinal gravitino behaves exactly as the goldstino fermion of supersymmetry breaking for energies and momenta higher than the gravitino mass.

\noindent In analogy with the case of the massive gauge boson, 
note that in the small mass limit the time-dependence and therefore the production of transverse polarizations of the Rarita--Schwinger field is suppressed. Likewise, the gravitational production of the longitudinal gravitino components is enhanced compared to the transverse components, since the couplings of the former are suppressed by the Planck mass, while the couplings of the latter are suppressed  by  the supersymmetry breaking scale. Therefore, the 
longitudinal components will be more copiously produced after inflation, setting constraints on the supersymmetry breaking scale, on the scale of inflation, and on the following thermal history.

Using  the results obtained in \cite{Bonnefoy:2022rcw}, we can study the lagrangian (\ref{eqn:L_theta}) of the longitudinal gravitino from the viewpoint of the equivalence theorem and connect it with the Goldstino lagrangian computed in \cite{Bonnefoy:2022rcw}. In this framework, the starting theory (\ref{eqn:L_psi}) is obtained from coupling the pure supergravity multiplet to two chiral superfields $\bm{S}$ and $\bm{\Phi}$ subject to the following superfield constraints \cite{Ferrara:2015tyn}
\begin{equation}
	\bm{S}^2=0 \ , \qquad \qquad \bm{S}\(\bm{\Phi}-\bar{\bm{\Phi}}\)=0 \ .
\label{eqn:sugra_constr}
\end{equation}

\noindent The first constraint above is the standard nilpotent constraint of the goldstino multiplet, whereas the second is the so called orthogonal constraint \cite{Komargodski:2009rz}. The constraints reduce the degrees of freedom associated with the chiral superfields $\bm{S}$ and $\bm{\Phi}$ to a real scalar field $\phi$ and a spin-$1/2$ fermion $G$, which can be absorbed in the unitary gauge in the longitudinal component of the gravitino. In the rigid limit, $G$ is identified with the Goldstino. The lagrangian (\ref{eqn:L_psi}) is then obtained from the following supergravity K\"ahler potential and superpotential \cite{Ferrara:2015tyn}
\begin{equation}
	K=\bm{S}\bar{\bm{S}}-\frac{1}{4}\(\bm{\Phi}-\bar{\bm{\Phi}}\)^2 \ , \qquad \qquad W=f(\bm{\Phi})\bm{S}+g(\bm{\Phi}) \ ,
\label{eqn:sugra_pot}
\end{equation}

\noindent with $f$ and $g$ that can be taken to be real without losing generality. The quadratic Goldstino lagrangian associated to the model (\ref{eqn:sugra_pot}) has been computed in \cite{Bonnefoy:2022rcw}. Considering a homogeneous background solution\footnote{In the following we use cosmic time $t$ instead of the conformal time $\tau$ used so far. The derivative with respect to the cosmic time will be denoted with the symbol $\dot{}$, while the symbol $'$ used until now to denote the derivative with respect to conformal time will denote here the field derivatives of the functions $f$ and $g$.}  for the real scalar $\phi=\phi(t)$ and working in momentum space and with our current conventions, this lagrangian is\footnote{In (\ref{eqn:L_G}) only the quadratic sector of the Goldstino lagrangian of \cite{Bonnefoy:2022rcw} is reported, being the one relevant for the equivalence theorem.}
\begin{equation}
	\begin{aligned}
		\mathcal{L}_G=&-\(1+\frac{\dot{\phi}^2}{2f^2}\)\frac{i}{2}\partial_0 G_{\bm{k}}\sigma^0\bar{G}_{\bm{k}}-\(1-\frac{\dot{\phi}^2}{2f^2}\)\frac{1}{2}G_{\bm{k}}\(\vk\cdot\vec{\sigma}\)\bar{G}_{\bm{k}} + \\
		& + i\frac{g'\,\dot{\phi}}{f^2}G_{\bm{k}}\(\vk\cdot\vec{\sigma}\)\bar{\sigma}^0G_{-\bm{k}}+h.c.\ ,
	\end{aligned}
\label{eqn:L_G}
\end{equation}

\noindent in which $G_{\bm{k}}$ is the Goldstino field in momentum space and we introduced the quantity $\vk\cdot\vec{\sigma}$ in analogy with eq. (\ref{eqn:k_contr}).

Our goal is to check that in the limit of high momenta the lagrangian for the longitudinal gravitino (\ref{eqn:L_theta_0}) (namely the one before the canonical normalisation (\ref{eqn:theta_cn})) matches  the rigid supersymmetry Goldstino lagrangian (\ref{eqn:L_G}). To this purpose, we have to use field equations in the FLRW background sourced  by the dynamics of the scalar field $\phi$. Its energy-momentum tensor is
\begin{equation}
\begin{aligned}
	T^{(\phi)}_{\mu\nu}=&\partial_\mu\phi\partial_\nu\phi-g_{\mu\nu}\[\frac{1}{2}\(\partial\phi\)^2-V(\phi)\]=\\
	=&\delta^0_\mu\delta^0_\nu\[\frac{\dot{\phi}^2}{2}+V(\phi)\]+a^2\delta_{ij}\delta^i_\mu\delta^j_\nu\[\frac{\dot{\phi}^2}{2}-V(\phi)\] \ .
\end{aligned}
\label{eqn:T_scalar}
\end{equation}

\noindent The supergravity scalar potential in the constrained superfield setup (\ref{eqn:sugra_constr}) is given by
\begin{equation}
	V(\phi)=f^2(\phi)-3g^2(\phi) \ ,
\end{equation}

\noindent whereas the gravitino mass is given by $m(\phi)=g(\phi)$. The FLRW background is then described by the following Einstein equations:
\begin{equation}
\begin{aligned}
	3\(H^2+m^2\)=\alpha=&\frac{\dot{\phi}^2}{2}+f^2 \ , \\
	\(\frac{R}{3}+H^2-3m^2\)=&\frac{\dot{\phi}^2}{2}-f^2 \ ,
\end{aligned}
\label{eqn:EE_bkgr}
\end{equation}

\noindent and the function $C(\tau)$ defined in (\ref{eqn:const_coeff}) becomes
\begin{equation}
	C(t) \ = \ \frac{\frac{\dot{\phi}^2}{2}-f^2}{\frac{\dot{\phi}^2}{2}+f^2}+i\frac{2g'\dot{\phi}}{\frac{\dot{\phi}^2}{2}+f^2}\fgamma^0 \ .
\label{eqn:C_bkgr}
\end{equation}

In two-components notation and in momentum space, the spinor $\theta$  takes the form\footnote{The $-\bm{k}$-dependence of the $\bar{\chi}$-component in (\ref{eqn:theta_2c_k}) is a consequence of the Majorana condition $\btheta=\theta^\textup{T}\mathcal{C}$ in momentum space.}
\begin{equation}
	\theta(t,\vk)=\begin{pmatrix}\chi_{\bm{k}}(t)\\\bar{\chi}_{-\bm{k}}(t)\end{pmatrix} \ .
\label{eqn:theta_2c_k}
\end{equation}

\noindent Making use of (\ref{eqn:4to2_gamma}), (\ref{eqn:EE_bkgr}) and (\ref{eqn:C_bkgr}), one finds 
\begin{equation}
\begin{aligned}
	\mathcal{L}_\theta=&\frac{\alpha a^2}{4\mk}\[i\btheta\fgamma^0\partial_0\theta-\btheta\kg C\theta +\frac{am}{2}\btheta\theta+\frac{3}{2}\(amC_\textup{R}+aHC_\textup{I}\)\btheta\theta\] = \\
	= &\frac{a^2f^2}{2\vk^2}\left\{-\(1+\frac{\dot{\phi}^2}{2f^2}\)\frac{i}{2}\partial_0\chi_{\bm{k}}\sigma^0\bar{\chi}_{\bm{k}}-\(1-\frac{\dot{\phi}^2}{2f^2}\)\frac{1}{2}\chi_{\bm{k}}\(\vk\cdot\vec{\sigma}\)\bar{\chi}_{\bm{k}}\right.+\\
	&\left. +\frac{g'\,\dot{\phi}}{f^2}\chi_{\bm{k}}\(\vk\cdot\vec{\sigma}\)\bar{\sigma}^0\chi_{-\bm{k}}+\frac{1}{2}\[\(\frac{\dot{\phi}^2}{f^2}-1\)am+\(\frac{3g'\,\dot{\phi}}{f^2}\)aH\]\chi_{\bm{k}}\chi_{-\bm{k}}+h.c.\right\} \ .
\end{aligned}
\label{eqn:L_theta_bkgr}
\end{equation}

We thus see that in the high momenta limit $k\gg am$ and $k\gg aH$, the longitudinal gravitino lagrangian (\ref{eqn:L_theta_0}),(\ref{eqn:L_theta_bkgr}) matches the Goldstino one (\ref{eqn:L_G}) with the identification
\begin{equation}
    \chi_{\bm{k}}=\frac{2k M_\textup{P}}{af}G_{\bm{k}} \ ,
\label{eqn:G_theta_id}
\end{equation}

\noindent where we reinstated the Planck mass $M_\textup{P}$ for dimensional reasons. This identification suggests the following prescription for the equivalence theorem:
\begin{equation}
	\psi_\mu\longrightarrow\frac{M_\textup{P}}{f}\partial_\mu G \ . \label{eq:equiv}
\end{equation}

\noindent Notice that during inflation the flat space relation $f= \sqrt{3} m M_P$ is not valid. Therefore only in flat space the factor $M_P/f$ in eq. (\ref{eq:equiv}) can be replaced by  $1/({{\sqrt 3} m}) $. 

One could legitimately wonder how to compute the hamiltonian and the energy operator from the goldstino action displayed in eq. (\ref{eqn:L_G}). 
The straightforward procedure of computing the hamiltonian starting from the  Goldstino field $G$ and its canonically conjugate momentum leads to a hamiltonian equal, in the high-momentum limit where the equivalence theorem applies, to the longitudinal gravitino instantaneous hamiltonian $\H_\vartheta$ (\ref{eqn:H_theta}). As already stressed, this is not the same as the energy operator of the longitudinal gravitino in supergravity $\E_\vartheta = T^0{}_0 (\vartheta)$. We do not currently know how to obtain this energy operator starting from the goldstino action (\ref{eqn:L_G}). Actually, the puzzle does not seem to be restricted only to our nonlinear supergravity model with orthogonal constraint (\ref{eqn:sugra_constr}). Indeed, in standard supergravity with the gravity multiplet and only one chiral multiplet containing the inflaton and the inflatino, in the unitary gauge the inflatino becomes the longitudinal gravitino component \cite{Kallosh:1999jj, Kallosh:2000ve}. The main difference between our previously considered constrained model and this more standard supergravity case is that the scalar field is now complex, containing an additional scalar together with the inflaton. It is an open and interesting question in this case as well to find the correct energy operator $\E_\vartheta = T^0{}_0 (\vartheta)$ starting from the effective goldstino action.

\section{Further comments and prospects}\label{sec:conclusion}

One main motivation for our work was the gravitino production in very specific supergravity settings discussed in \cite{Hasegawa:2017hgd,Kolb:2021xfn,Kolb:2021nob,Dudas:2021njv}, in the case where the sound speed for the longitudinal component vanishes at some point(s) during the cosmological evolution. While it was suggested in \cite{Kolb:2021xfn,Kolb:2021nob} that this puzzling result may signal an intrinsic incompatibility between certain effective theories of supergravity and a UV-completion into quantum gravity, there are reasons to believe that such behaviour will be modified in the UV. Firstly, as already anticipated in \cite{Kolb:2021xfn, Kolb:2021nob}, the effective field theory in this case should break down before the Planck scale. This happens when the kinetic energy of the scalar (inflaton) field equals the energy provided by supersymmetry breaking ${\dot \phi}^2/2 = f^2$. The goldstino action (\ref{eqn:L_G}) worked out in \cite{Bonnefoy:2022rcw} strongly suggests that in this case the effective field theory breaks down, since the effects of higher-derivative operators are not anymore small. Actually the cutoff $\Lambda$ of supergravity theories with nonlinear supersymmetry is not the Planck mass, but much lower and determined by the supersymmetry breaking scale $\Lambda \sim \sqrt{f}$ \cite{unitarity}. Therefore the limit of zero sound speed in such models arises precisely when the inflaton kinetic energy reaches the cutoff of the effective field theory.  

On the other hand, it is still interesting to investigate the backreaction on the geometry of gravitino production in the case of a small non-vanishing value of the sound speed. 
The gravitino energy-momentum tensor (\ref{eqn:emt_full}) that we computed in our paper is the first ingredient needed for this purpose. The puzzles that we raised, namely the mismatch of the Fock space of the energy operator with respect to the Fock space of the instantaneous hamiltonian, even for high-momenta, and the computation of the energy operator in the goldstino action in the regime of validity of the equivalence theorem, clearly deserve further dedicated studies. 
It is also interesting that the sound speed does not enter directly the energy operator for the longitudinal gravitino. This result seems valid also in usual supergravity with one chiral multiplet (the inflaton), where the sound speed equals one, but the energy operator is more complicated. Moreover, at first sight the energy operator does not feature anything special in the case of zero sound speed, in the models based on non-linear supersymmetry where this can happen. A full analysis of the backreaction involves however not only a discussion of the zero-point energy, but a full understanding of the renormalization of the energy operator. We plan to return to these issues in the near future.  

\section*{\sc Acknowledgments}
\vskip 12pt
We thank  Bohdan Grzadkowski, Yann Mambrini, Hans Peter Nilles and Anna Socha  for stimulating discussions. GC and ED are grateful to CERN-TH for the kind hospitality and GC is grateful to the Univ. of Padova for hospitality and support while this work was in progress. MP is supported by Istituto
Nazionale di Fisica Nucleare (INFN) through the Theoretical Astroparticle Physics (TAsP)
project. 

\newpage
\appendix


\section{Algebra of the helicity eigenstate decomposition}\label{app:hed_algebra}

We list here in more detail the algebraic equations and properties of the helicity eigenstate decomposition of the gravitino employed in Section \ref{section:spin_32}. The indices contraction rules and notation are
\begin{equation}
\begin{aligned} 
    \vec{k}\cdot\vec{x}=&\sum_{i=1}^{3}k^ix^i=-k^ix_i \ , & 
\vfgamma\cdot\vpsi_k=&\sum_{i=1}^3\fgamma^i\psi_{i,k}=\fgamma^i\psi_{i,k} \ , \\
	\vk\cdot\vpsi_k=&\sum_{i=1}^3k^i\psi_{i,k}=k^i\psi_{i,k} \ , &
	\vec{k}\cdot\vfgamma=&\sum_{i=1}^{3}k^i\fgamma^i=-k_i\fgamma^i \ ,
\end{aligned}
\label{eqn:k_contr}
\end{equation}

\noindent and the following identities hold:
\begin{align}
    \kg\kg=-\mk& \ , & \kg\fgamma_i\kg = &\mk\,\fgamma_i-2k_i\kg \ .
\end{align}

\noindent The projectors introduced in (\ref{eqn:psi_proj}) satisfy the following algebra:
\begin{equation}
\begin{aligned}
    &\fgamma^i\left(P_\gamma\right)_i=1 \ , &&& &\fgamma^i\left(P_k\right)_i=0 \ , \\
	&k^i\left(P_\gamma\right)_i=0 \ , &&&    &k^i\left(P_k\right)_i=1 \ , \\
	&\left\{\left(P_\gamma\right)_i,\fgamma_0\right\}=0 \ ,	&&&		&\left\{\left(P_\gamma\right)_i,\fgamma_j\right\}=\eta_{ij}+\frac{k_ik_j}{\mk} \ , \\
	&\left[\left(P_\gamma\right)_k,\fgamma_0\right]=0 \ ,	&&&		&\left[\left(P_\gamma\right)_k,\fgamma_j\right]=\frac{1}{\mk}\left[\fgamma_i k_j+\eta_{ij}\kg\right] \ , \\
    &\left(P_\gamma\right)_i^\dagger=-\left(P_\gamma\right)_i \ , &&& &\left(P_k\right)_i^\dagger=-\frac{1}{2\mk}\left[3k_i+\left(\vk\cdot\vfgamma\right)\fgamma_i\right] \ , \\
    &\left(P_\gamma\right)^i\left(P_\gamma\right)_i=\frac{1}{2} \ , &&& &\left(P_\gamma\right)^i\left(P_k\right)_i=-\frac{1}{2\mk}\kg \ ,\\
    &\left(P_k\right)^i\left(P_k\right)_i=-\frac{1}{2\mk} \ , &&&   &\left(P_k\right)^i\left(P_\gamma\right)_i=\frac{1}{2\mk}\kg \ ,\\
    &\left(P_k\right)^{i\,\dagger}\left(P_k\right)_i=-\frac{3}{2\mk} \ , &&& &\left(P_k\right)^{i\,\dagger}\left(P_\gamma\right)_i=-\frac{1}{2\mk}\kg \ .
\end{aligned}
\label{eqn:proj_algebra}
\end{equation}

\noindent The operator $\hO_i$ introduced in (\ref{eqn:psi_dec})-(\ref{eqn:psi_dec_O}) satisfies instead the following identities:
\begin{align}
    \fgamma^0\hO_i^\dagger\fgamma^0=&\frac{1}{\mk}\left\{k_i\kg-\frac{1}{2}\left(am-iaH\fgamma^0\right)\left[3k_i+\kg\fgamma_i\right]\right\} \ , \\
    \kg\hO_i=&-\left\{k_i+\frac{1}{2}\left[\frac{k_i}{\mk}\kg+\fgamma_i\right]\left(am+iaH\fgamma^0\right)\right\} \ .
\end{align}

\section{Symmetry and covariant conservation of the gravitino energy-momentum tensor}\label{app:emt}

In this Appendix we derive the gravitino energy-momentum tensor. We also explicitly show that, once evaluated on-shell, this tensor is symmetric and covariantly conserved. In the case of the Rarita--Schwinger spin-$\nicefrac{3}{2}$ field that we considered, the variation of the action (\ref{eqn:L_psi}) with respect to the veirbein gives\footnote{In equation (\ref{eqn:T_non_sym}), the EoM (\ref{eqn:gr_eom_1}) have already been applied once.}:
\begin{equation}
	\begin{aligned}
		T^{\mu\nu}=&\frac{i}{2}\bPsi_\rho\gamma^{\rho \mu \alpha}\({\nabla}_\alpha\Psi^\nu-{\nabla}^\nu\Psi_\alpha\)-\frac{m}{2}\bPsi_\rho\gamma^{\rho\mu}\Psi^\nu\\
		&+\frac{i}{4}\nabla_\rho\(\bPsi^\rho\gamma^\mu\Psi^\nu+\bPsi^\rho\gamma^\nu\Psi^\mu-\bPsi^\mu\gamma^\rho\Psi^\nu\)\\
		&+\frac{i}{2}\nabla^\nu\left(\bPsi^\mu\gamma^\rho\Psi_\rho\)+\frac{i}{2}g^{\mu\nu}\,\nabla_\rho\left(\bPsi_\alpha\gamma^\alpha\Psi^\rho\right) \ .
	\end{aligned}
\label{eqn:T_non_sym}
\end{equation}

\noindent Using then the EoM (\ref{eqn:gr_eom_1})-(\ref{eqn:gr_eom_2}), together with the identities
\begin{align}
	\epsilon^{\lambda\mu\nu\rho}\epsilon_{\lambda\alpha\beta\gamma}=-3!\,\delta^{[\mu}_{\alpha}\delta^{\nu}_{\beta}\delta^{\rho]}_{\gamma} \ , && \gamma^{\mu\nu\rho}=i  \epsilon^{\lambda\mu\nu\rho}\gamma_5\gamma_\lambda \ , 
\label{eqn:ids_g}
\end{align}

\noindent one can show that the tensor in (\ref{eqn:T_non_sym}) has vanishing antisymmetric part:
\begin{equation}
\begin{aligned}
    T^{[\mu\nu]}=&\frac{i}{2}\bPsi_\rho\gamma^{\rho [\mu| \alpha}\({\nabla}_\alpha\Psi^{|\nu]}-{\nabla}^{|\nu]}\Psi_\alpha\)-\frac{m}{2}\bPsi_\rho\gamma^{\rho[\mu|}\Psi^{|\nu]}\\
    &+\frac{i}{2}\nabla^{[\nu}\(\bPsi^{\mu]}\gamma^\rho\Psi_\rho\)-\frac{i}{4}\nabla_\rho\(\bPsi^{[\mu|}\gamma^\rho\Psi^{|\nu]}\)=\\
    =&\frac{i}{2}\bPsi_\rho\gamma^{\rho[\mu}\[\cancel{\nabla}\psi^{\nu]}-\nabla^{\nu]}\(\gamma^\rho\Psi_\rho\)\]+\frac{i}{2}\bPsi^\alpha\gamma^{[\mu}\(\nabla_\alpha\Psi^{\nu]}-\nabla^{\nu]}\Psi_\alpha\)-\frac{m}{2}\bPsi_\rho\gamma^{\rho[\mu|}\Psi^{|\nu]}\\
    &+\frac{i}{2}\bPsi_\rho\gamma^\rho\nabla^{[\nu}\Psi^{\mu]}-\frac{i}{2}\bPsi^{[\mu}\[\cancel{\nabla}\Psi^{\nu]}-\nabla^{\nu]}\(\gamma^\rho\Psi_\rho\)\]\overset{(\ref{eqn:gr_eom_2})}{=}\\
    =&\frac{m}{2}\bPsi_\rho\gamma^{\rho[\mu}\Psi^{\nu]}+\frac{3}{2}i\bPsi_\rho\gamma^{[\rho}\nabla^\nu\Psi^{\mu]}=\\
    =&\frac{m}{2}\bPsi_\rho\gamma^{\rho[\mu}\Psi^{\nu]}+\frac{3}{2}i\delta^{[\rho}_{\alpha}\delta^{\nu}_{\beta}\delta^{\mu]}_{\gamma}\bPsi_\rho\gamma^\alpha\nabla^\beta\Psi^\gamma\overset{(\ref{eqn:ids_g})}{=}\\
	=&\frac{m}{2}\bPsi_\rho\gamma^{\rho[\mu}\Psi^{\nu]}-\frac{1}{4}\epsilon^{\rho\nu\mu}{}_{\lambda}\bPsi_\rho\gamma_5\[i\epsilon^{\alpha\lambda\beta\gamma}\gamma_5\gamma_\alpha\]\nabla_\beta\Psi_\gamma\overset{(\ref{eqn:ids_g})}{=}\\
=&\frac{m}{2}\bPsi_\rho\gamma^{\rho[\mu}\Psi^{\nu]}-\frac{1}{4}\epsilon^{\rho\nu\mu}{}_{\lambda}\bPsi_\rho\gamma_5\[\gamma^{\lambda\beta\gamma}\nabla_\beta\Psi_\gamma\]\overset{(\ref{eqn:gr_eom_1})}{=}\\
 =&\frac{m}{2}\bPsi_\rho\gamma^{\rho[\mu}\Psi^{\nu]}+\frac{m}{4}\bPsi_\rho\[i\epsilon^{\lambda\rho\nu\mu}\gamma_5\gamma_{\lambda}\]\gamma^{\alpha}\Psi_\alpha\overset{(\ref{eqn:ids_g})}{=}\\
 =&\frac{m}{2}\bPsi_\rho\gamma^{\rho[\mu}\Psi^{\nu]}+\frac{m}{4}\bPsi_\rho\gamma^{\rho\nu\mu}\gamma^{\alpha}\Psi_\alpha=\frac{m}{2}\bPsi_\rho\gamma^{\rho[\mu}\Psi^{\nu]}-\frac{m}{2}\bPsi_\rho\gamma^{\rho[\mu}\Psi^{\nu]}=0 \ .
\end{aligned}
\end{equation}
This allows us to recast eq.~(\ref{eqn:T_non_sym}) in the manifestly symmetric form (\ref{eqn:emt_full}). 

The proof of the covariant conservation goes along similar lines and requires the constraints (\ref{eqn:const_1})-(\ref{eqn:const_2}) (a direct consequence of the EoM) and the following identities:
\begin{equation}
\begin{aligned}
    R^\mu_{[\nu\rho\sigma]}=&\frac{1}{3}\[R^\mu_{\nu\rho\sigma}+R^\mu_{\rho\sigma\nu}+R^\mu_{\sigma\nu\rho}\]=0 \ , \\
    R_{\mu\nu\alpha\beta}\gamma^{\mu\nu\rho}\gamma^{\alpha\beta}=&4G^\rho_\alpha\gamma^\alpha \ ,\\
    \left[\nabla_\mu,\nabla_\nu\right]\Psi^\rho=&\frac{1}{4}R_{\mu\nu\,ab}\gamma^{ab}\Psi^\rho+R^\rho{}_{\sigma\mu\nu}\Psi^\sigma \ , \\
    \[\nabla_\mu,\nabla_\nu\]V^\rho=&R^\rho{}_{\sigma\mu\nu}V^\sigma \ .
\end{aligned}
\label{eqn:ids_R}
\end{equation}

\noindent We also notice that, since its antisymmetric part is identically vanishing, we can perform this computation on the bare tensor (\ref{eqn:T_non_sym}), having a more convenient form. We start from the first line of (\ref{eqn:T_non_sym}), for which we have
\begin{equation}
\begin{aligned}
    \frac{i}{2}\nabla_\mu&\[\bPsi_\rho\gamma^{\rho\mu\alpha}\(\nabla_\alpha\Psi^\nu-\nabla^\nu\Psi_\alpha\)-m\bPsi_\rho\gamma^{\rho\mu}\Psi^\nu\]=\\ 
	&=\frac{i}{2}\bPsi_\rho\lnabla_\mu\gamma^{\rho\mu\alpha}\(\nabla_\alpha\Psi^\nu-\nabla^\nu\Psi_\alpha\)+\frac{i}{2}\bPsi_\rho\gamma^{\rho\mu\alpha}\nabla_\mu\nabla_\alpha\Psi^\nu-\frac{i}{2}\bPsi_\rho\gamma^{\rho\mu\alpha}\nabla_\mu\nabla^\nu\Psi_\alpha\\
	&\quad -\frac{m}{2}\bPsi_\rho\lnabla_\mu\gamma^{\rho\mu}\Psi^\nu-\frac{m}{2}\bPsi_\rho\gamma^{\rho\mu}\nabla_\mu\Psi^\nu-\frac{i}{2}\(\partial_\mu m\)\bPsi_\rho\gamma^{\rho\mu}\Psi^\nu\overset{(\ref{eqn:gr_eom_1}), (\ref{eqn:const_1}),(\ref{eqn:ids_R})}{=}\\
	&=\frac{3}{4}im^2\bPsi_\rho\gamma^\rho\Psi^\nu-\frac{i}{2}\(\partial_\mu m\)\bPsi_\rho\gamma^{\rho\mu}\Psi^\nu+\frac{i}{4}G^\rho\,_\alpha\bPsi_\rho\gamma^\alpha\Psi^\nu\\
    &\quad +\frac{i}{16}R^\nu_{\,\mu\lambda\tau}\bPsi_\rho\left\{\gamma^{\rho\mu\alpha},\gamma^{\lambda\tau}\right\}\Psi_\alpha\overset{(\ref{eqn:const_2})}{=}
	\frac{i}{4}R^\nu_{\,\rho\alpha\beta}\bPsi^{\alpha}\gamma^\rho\Psi^\beta+\frac{i}{2}R^\nu_{\,\rho}\bPsi_\alpha\gamma^\alpha\Psi^\rho \ .
\end{aligned}
\label{eqn:cov_cons_kin}
\end{equation}

\noindent All the remaining terms are in the form of two covariant derivatives acting on a term of the type $\bPsi^\mu\gamma^\nu\Psi^\rho\equiv V^{\mu\nu\rho}$. From the point of view of general diffeomorphisms, this combination transforms as a tensor in the covariant indices, i.e. its covariant derivative does not carry anymore the terms associated to the spinorial nature of the gravitino. This allows to simplify the computation of the remaining terms because we only need the last equation of (\ref{eqn:ids_R}). Thus, we have
\begin{align}
&\begin{aligned}
	\bullet \quad \frac{i}{4}\nabla_\mu\nabla_\rho&\[\bPsi^\rho\gamma^\nu\Psi^\mu\]=\frac{i}{8}\[\nabla_\mu,\nabla_\rho\]V^{\rho\nu\mu}=\\
	=&\frac{i}{8}\(R^{\rho}_{\sigma\mu\rho}V^{\sigma\nu\mu}+R^{\mu}_{\sigma\mu\rho}V^{\rho\nu\sigma}+R^{\nu}_{\sigma\mu\rho}V^{\rho\sigma\mu}\)=-\frac{i}{8}R^{\nu}_{\rho\alpha\beta}\bPsi^\alpha\gamma^\rho\Psi^\beta \ ,
	\end{aligned}\\\notag\\
&\begin{aligned}
	\bullet \quad \frac{i}{4}\nabla_\mu\nabla_\rho&\[\bPsi^\rho\gamma^\mu\Psi^\nu-\bPsi^\mu\gamma^\rho\Psi^\nu\]=\frac{i}{4}\[\nabla_\mu,\nabla_\rho\]V^{\rho\mu\nu}=\\
	=&\frac{i}{4}\(R^{\rho}_{\sigma\mu\rho}V^{\sigma\mu\nu}+R^{\mu}_{\sigma\mu\rho}V^{\rho\sigma\nu}+R^{\nu}_{\sigma\mu\rho}V^{\rho\mu\sigma}\)=\frac{i}{4}R^\nu_{\beta\rho\alpha}\bPsi^\alpha\gamma^\rho\Psi^\beta \ ,
\end{aligned}\\\notag\\
&\begin{aligned}
	\bullet \quad \frac{1}{2}\nabla_\mu&\[ig^{\mu\nu}\nabla_{\rho}\(\bPsi_\alpha\gamma^\alpha\Psi^\rho\)+i\nabla^\nu\(\bPsi^\mu\gamma^\rho\Psi_\rho\)\]=\\
	&=\frac{i}{2}\[\nabla^\nu,\nabla_\mu\]\(\bPsi_\rho\gamma^\rho\Psi^\mu\)V^\mu=-\frac{i}{2}R^\nu_\rho\bPsi_\alpha\gamma^\alpha\Psi^\rho \ .
	\end{aligned} \label{eqn:cov_cons_vec}
\end{align}

\noindent Therefore, summing up equations (\ref{eqn:cov_cons_kin})-(\ref{eqn:cov_cons_vec}) and applying again the Bianchi identity of the Riemann tensor in (\ref{eqn:ids_R}), we indeed find that the energy-momentum tensor (\ref{eqn:emt_full}) is covariantly conserved
\begin{equation}
    \nabla_\mu T^{\mu\nu}=0 \ .
\end{equation}


\end{document}